\documentclass[reprint,aps,pre,onecolumn,notitlepage]{revtex4-1}

\usepackage[T1]{fontenc} %
\usepackage[utf8]{inputenc} %
\usepackage[english]{babel}		%
\usepackage{amsmath}	%
\usepackage{amssymb}
\usepackage{graphicx} %
\usepackage{hyperref}
\usepackage[]{cleveref}
\usepackage{siunitx}	%
\usepackage{tabu}
\usepackage{makecell}

\Crefname{equation}{Eq.}{Eqs.}

\newcommand{\be}{\begin{equation}}
\newcommand{\ee}{\end{equation}}

\newcommand{\nn}{\nonumber}
\newcommand{\lp}{\left(}
\newcommand{\rp}{\right)}
\newcommand{\lbk}{\left[}
\newcommand{\rbk}{\right]}
\newcommand{\di}{\mathrm{d}}

\begin{document}
	
\title{Analytical cell size distribution: lineage-population bias and parameter inference}
\author{Arthur Genthon}
\affiliation{Gulliver UMR CNRS 7083, ESPCI Paris, Université PSL, 75005 Paris, France}
\email{arthur.genthon@espci.fr}

\begin{abstract} 

We derive analytical steady-state cell size distributions for size-controlled cells in single-lineage experiments, such as the mother machine, which are fundamentally different from batch cultures where populations of cells grow freely.
For exponential single-cell growth, characterizing most bacteria, the lineage-population bias is obtained explicitly. In addition, if volume is evenly split between the daughter cells at division, we show that cells are on average smaller in populations than in lineages. 
For more general power-law growth rates and deterministic volume partitioning, both symmetric and asymmetric, we derive the exact lineage distribution. This solution is in good agreement with Escherichia coli mother machine data, and can be used to infer cell cycle parameters such as the strength of the size control and the asymmetry of the division. 
When introducing stochastic volume partitioning, we derive the large-size and small-size tails of the lineage distribution, and show that the lineage-population bias only depends on the single-cell growth rate. These asymptotic behaviors are extended to the adder model of cell size control. 
When considering noisy single-cell growth rate, we derive the large-size lineage and population distributions. Finally, we show that introducing noise, either on the volume partitioning or on the single-cell growth rate, can cancel the lineage-population bias.

\end{abstract}

\maketitle

\section{Introduction}

In the past decade, a large amount of single-cell data has been obtained thanks to microfluidic devices, such as the mother machine \cite{wang_robust_2010}. In these experiments, time-lapse video-microscopy allows to follow single-cell lineages over many generations with great precision, resulting in large and reliable statistics. 
Several questions are then naturally raised: How can we use these data to infer the laws of cell growth and division? Can we learn population-level properties from single-lineage measurements? Are single-lineage statistics different from population statistics obtained in batch cultures? 

The first two questions received recent attention. For example, Jia et al. proposed a method to infer single-cell parameters from size distributions, both for bacteria \cite{jia_cell_2021} and yeasts \cite{jia_characterizing_2022}.  
Also, single-lineage statistics on the number of divisions can in principle be used to estimate the population growth rate, with which the population would grow in a batch culture \cite{levien_large_2020,genthon_fluctuation_2020,pigolotti_generalized_2021}.

The answer to the third question is yes, and quantifying the differences between these two perspectives is fundamental in order to compare and analyze the different sets of data. This problem can be traced back to Powell's 1956 seminal work \cite{powell_growth_1956}, in which he showed that for age-structured populations in exponential growth, the lineage and population distributions of generation times (time elapsed between birth and division) are different. This difference is understood as cells that divided more than average lead to subpopulations of offsprings that are over-represented in the population, while no such selection is present in single-lineage experiments. Powell's results have been generalized in several directions since \cite{genthon_fluctuation_2020,levien_interplay_2020,nakashima_lineage_2020}.
The problem was recently reformulated by Nozoe et al. \cite{nozoe_inferring_2017} in terms of two different samplings of lineages in a population tree, and their framework has been used to further investigate the lineage-population bias \cite{thomas_making_2017,thomas_analysis_2018}, its consequences on selection strength \cite{genthon_universal_2021}, and has been analyzed through the lens of stochastic thermodynamics \cite{garcia-garcia_linking_2019,genthon_fluctuation_2020}. 
The lineage-population bias for any cell trait can be studied using the notion of fitness landscape \cite{nozoe_inferring_2017}, which quantifies the correlations between the value of the trait and the number of divisions undergone by the cell. We showed in some simple cases that for age and size, the population distributions of these traits are biased toward small values as compared to the lineage statistics \cite{genthon_fluctuation_2020}, because cells that divided a lot are on average smaller and younger.
Another route to investigate this bias, that we take in this article, is to solve independently the population and lineage equations.
\begin{figure*}
	\includegraphics[width=0.7\linewidth]{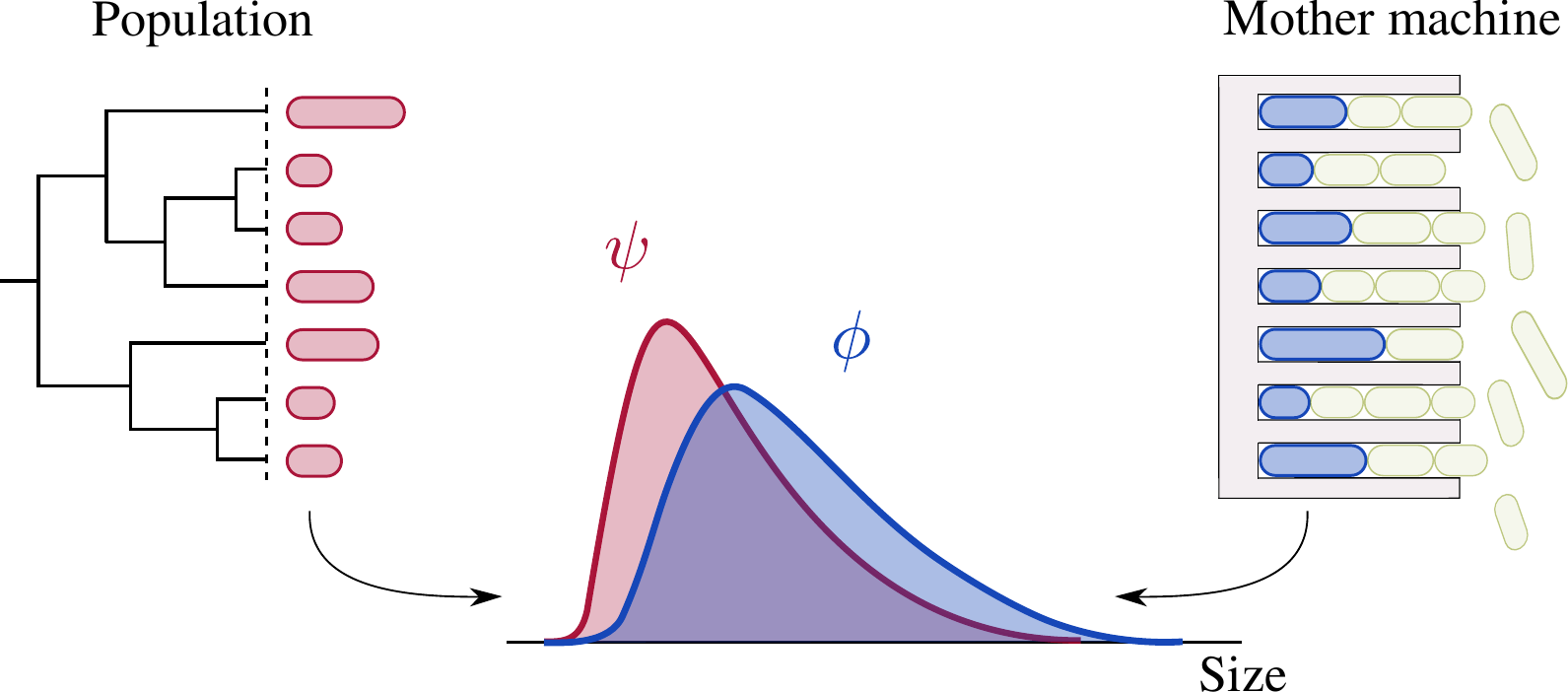}
	\caption{Snapshot cell-size distributions for two different experimental setups. The population distribution $\psi$ (red) is computed by uniformly sampling cell sizes in a freely growing population in a batch culture, and the lineage distribution $\phi$ (blue) is obtained by uniformly sampling the sizes of the constant number of mother cells (in blue) at the bottoms of each micro-channel in a mother machine device.}
	\label{fig_lin_vs_pop}
\end{figure*}

In this article, we focus on cell size which offers many insights on the laws of growth and division, and because the size framework also applies to molecular-level quantities, such as a number of proteins or mRNA, which also grow within the cell cycle and are split between the daughter cells at division. Lineage-population biases have been recently derived at the level of the first moments of the size distribution \cite{totis_population-based_2021}, and of the distribution of size at birth \cite{thomas_analysis_2018} in some particular cases; however general understanding of the bias at the level of distributions, as illustrated on \cref{fig_lin_vs_pop}, is lacking. 
In the mathematical literature, the growth-fragmentation equation modeling the time evolution of the population cell-size distribution has been largely analyzed \cite{michel_existence_2006,doumic_jauffret_eigenelements_2010,balague_fine_2013}. 
Surprisingly, these analyses have not been applied to lineage statistics. 

The fluctuations in cell growth are of central importance for cell size statistics. Variability in single cell growth has been mainly modeled as a Markov process, where the single cell growth rate changes from one cycle to the next one, but remains constant inside each cycle, so that the growth of a single cell is deterministic \cite{doumic_statistical_2015,garcia-garcia_linking_2019}. This kind of modeling accounts for cell-to-cell variability, which affects the population growth rate \cite{olivier_how_2017}, either increasing or decreasing it depending on mother-daughter correlations \cite{lin_single-cell_2020}. 
On the other hand, experimental cycles exhibit fluctuations around the exponential trend, prompting us to describe single-cell growth as a random process with a diffusive term accounting for in-cycle variability. Experimental data on E. coli \cite{kiviet_stochasticity_2014} suggest that both in-cycle and cell-to-cell sources of variability are present. In this article, we focus on in-cycle variability in cell growth, which has been less investigated than cell-to-cell variability.

The goals of this article are multiple and are summarized as follows. After introducing the size-control model in \cref{sec_prelim}, we derive in \cref{sec_exact_lin} exact steady-state lineage cell-size distributions, in the case of deterministic volume partitioning, both symmetric and asymmetric. The lineage-population bias is then obtained by comparison with the mathematical literature on population. We also show that this result accounts for experimental data, and can be used for parameters inference. In \cref{sec_gen_kernel}, we introduce stochasticity in the partitioning of volume at division and seek large and small size asymptotic lineage distributions. From these distributions, we derive the lineage-population bias and investigate the role of the different rates of the model in this bias. Moreover, we show that these asymptotic behaviors are also valid for the adder mechanism, in which cell division is triggered the increment of volume since birth. Finally, in \cref{sec_noise} we introduce in-cycle noise in single-cell growth around the exponential trend, and show how it modifies the large-size behavior and the lineage-population bias.

\section{Preliminaries}
\label{sec_prelim}

\subsection{Model, definitions and hypotheses}

The expected number $n(x,t)$ of cells of size $x$ at time $t$ obeys the following population balance equation (PBE), also called diffusion-growth-fragmentation equation \cite{zaidi_probability_2016,tchouanti_well_2022}:
\begin{equation}
\label{eq_pop}
\partial_t n(x,t) = -\partial_x [\nu(x) n(x,t)] + \partial_{x^2} [D(x) n(x,t)] -r(x) n(x,t) + m \int \frac{\di x'}{x'} b(x/x') r(x') n(x',t) \,,
\end{equation}
supplemented with the `no-flux' boundary conditions at $x=0$ and $x=\infty$:
\begin{align}
0&=\lim\limits_{x \rightarrow 0 }\partial_{x} [D(x) n(x,t)] - \nu(x) n(x,t) \nn \\
&= \lim\limits_{x \rightarrow \infty}\partial_{x} [D(x) n(x,t)] - \nu(x) n(x,t) \,,
\end{align}
and the initial condition $n(x,t=0)=n_0(x)$.
In the right hand side, the first two terms account respectively for the deterministic part of the cell growth with rate $\nu(x)$, and for the stochastic part of the growth where $D(x)$ is the diffusion coefficient. In the corresponding Langevin representation of the cell cycle, size grows as $ \di x = \nu(x) \di t + \sqrt{2D(x)} \di W$ where $W$ is the Wiener process. 
This equation reduces to a growth-fragmentation equation when single-cell growth is deterministic ($D(x)=0$) \cite{michel_existence_2006}, to a diffusion-fragmentation equation when there is no single-cell growth ($\nu(x)=0$) \cite{laurencot_fragmentation_2021-1,laurencot_fragmentation_2021}, and to a fragmentation equation when $D(x)=\nu(x)=0$ \cite{cheng_scaling_1988}.

The third term describes the division of cells of size $x$ with a rate $r(x)$, and the last term accounts for the birth of $m$ new cells of size $x$ coming from the divisions of cells of sizes $x'$, through the partition kernel $b(x/x')$ which is the probability for a newborn cell to inherit the fraction $x/x'$ of its mother's volume at division. Most cells obey \textit{binary fission} corresponding to $m=2$. The partition kernel is normalized as $\int_{0}^{1}\di x \ b(x)=1$, and the conservation of volume at division imposes that $m \int_{0}^{1} \di x \ x b(x)=1$. Moreover, births of cells of size $0$ are impossible, so that we set $b(0)=b(1)=0$. The kernel $b$ is very general, and accounts for \textit{deterministic symmetric partition} observed in bacteria and fission yeast: $b(x)=\delta(x-1/m)$ where $\delta$ is the Dirac delta distribution; \textit{deterministic asymmetric partition} characterizing for example budding yeast: $b(x)=\sum_{i=1}^{m}\delta(x-1/\omega_i)/m$ with $\omega_i > 1$ and $\sum_{i=1}^{m} 1/\omega_i=1$; and \textit{stochastic partition} which can be modeled for example as a Beta distribution for size or as a Binomial distribution for protein segregation.

We recast this equation at the probability level by defining the frequency $\psi(x,t)=n(x,t)/N_t$ of cells of size $x$, with $N_t=\int \di x n(x,t)$ the total number of cells at time $t$:
\begin{equation}
\label{eq_pop_full}
\partial_t \psi(x,t) = -\partial_x [\nu(x) \psi(x,t)] + \partial_{x^2} [D(x) \psi(x,t)] - \lbk r(x) +\Lambda_t \rbk \psi(x,t) + m \int \frac{\di x'}{x'} b(x/x') r(x') \psi(x',t) \,,
\end{equation}
where $\Lambda_t=\partial_t N_t/N_t$ is the instantaneous population growth rate. Direct integration of this equation with respect to $x$ shows that this population growth rate is equal to the average division rate:
\begin{equation}
\label{eq_def_lam}
\Lambda_t=(m-1) \int \di x \ r(x) \psi(x,t) \,.
\end{equation}
We refer to this distribution $\psi$ as the population distribution, and draw the attention of the reader on the fact that adding a uniform ($x$-independent) death/dilution rate does not affect this population distribution \cite{basse_cell-growth_2004,levien_interplay_2020}. 
In particular, setting a dilution rate that perfectly balances the population growth allows to maintain constant populations, as it is done in chemostats, but does not affect the size-distribution.

The equation for the lineage distribution $\phi(x)$ is obtained from \cref{eq_pop_full} by tracking only $m=1$ daughter cell at division \cite{robert_division_2014}, thus maintaining the population constant with a null growth rate $\Lambda_t$:
\begin{equation}
\label{eq_lin_full}
\partial_t \phi(x,t) = -\partial_x [\nu(x) \phi(x,t)]  + \partial_{x^2} [D(x) \phi(x,t)] -r(x) \phi(x,t) + \int \frac{\di x'}{x'} b(x/x') r(x') \phi(x',t) \,.
\end{equation}
Importantly, we set $m=1$ without changing the partition kernel: even though we follow only one of the $m$ daughters at division, it still inherits $1/m$ of the mother volume (for deterministic symmetric partitioning) and not $1$. In this sense, this equation is not equivalent to the one obtained in the limit $m \rightarrow 1^{+}$ considered in \cite{hall_functional_1989}.
Note that, if the lineage distribution is defined as a snapshot distribution, it can be evaluated along a single lineage in time because of the ergodic principle. 

Through the article we use the moments of order $k$ of the distributions $b$, $\psi$ and $\phi$:
\begin{align}
L_{k}&= \int_{0}^{1} x^k b(x) \di x \\
M_{k}&= \int_{0}^{\infty} x^k \psi(x) \di x \\
N_{k}&= \int_{0}^{\infty} x^k \phi(x) \di x \,,
\end{align}
which are the Mellin transforms of these distributions (except for a simple shift of $1$ in the definition of exponent $k$).

Let us now detail the different assumptions made in the following sections. 
In \cref{sec_exact_lin}, we seek analytical solutions to the steady-state growth-fragmentation equation \cref{eq_lin_full} for deterministic single-cell growth ($D(x)=0$), with deterministic partitioning and power-law division and growth rates with respective powers $\alpha \geq0$ and $\beta \geq 0$:
\begin{align}
\label{eq_pl_r}
r(x)&=r x^{\alpha} \\
\label{eq_pl_nu}
\nu(x)&=\nu x^{\beta} \,.
\end{align}
This choice for the division rate can be justified theoretically \cite{nieto_unification_2020} and the power $\alpha$ is the strength of the size-control: in the limit $\alpha \rightarrow 0$ of weak control, cells of all sizes divide with the same rate, while in the limit $\alpha \rightarrow +\infty$ of strong control, cells divide deterministically when reaching size $x=1$. 
The power law for the growth rate includes the most common growth strategies, which we call \textit{linear growth} for $\beta=0$, and \textit{exponential growth} which characterizes most bacteria \cite{taheri-araghi_cell-size_2015} for $\beta=1$.
In the long-time limit, the exponential growth of the population with a rate $\Lambda=\lim\limits_{t \rightarrow \infty} \Lambda_t$ and the existence of a steady-state size distribution are ensured by stability conditions \cite{michel_existence_2006,doumic_jauffret_eigenelements_2010,balague_fine_2013}. Among them, 
\be
\label{eq_cond_bet_alp}
\alpha-\beta+1 >0
\ee
ensures that there is enough division to avoid cells of diverging sizes, and that there is enough growth to avoid cells of vanishing sizes \cite{michel_existence_2006}.
In \cref{sec_gen_kernel}, the hypothesis on deterministic partitioning is relaxed, and the assumptions \crefrange{eq_pl_r}{eq_cond_bet_alp} are replaced by similar ones in the limits of small and large sizes. 
Finally, in \cref{sec_noise} we introduce stochasticity in single-cell growth around the exponential trend ($\beta=1$), with a Gaussian noise on the growth rate itself \cite{alonso_modeling_2014}: $\di x = (\nu \di t+ \sqrt{2D} \di W) x$. This choice corresponds to a diffusion coefficient $D(x)=D x^2$ in \cref{eq_pop}, which we call \textit{multiplicative noise}.

\subsection{The special case of single-cell exponential growth}
\label{sec_exact_bias}

When cells grow exponentially with a rate $\nu(x)=\nu x$, two important steady-state results are derived for any kernel $b$ and any division rate $r$ (the power law assumption is not required).

First, the steady-state population growth matches the single cell growth rate \cite{hall_functional_1990}. This follows from the integration of \cref{eq_pop_full} after multiplication by $x$, and using the mass conservation property of kernel $b$:
\begin{equation}
\label{eq_dt_x_av}
\partial_t M_1 = (\nu-\Lambda_t) M_1 \,,
\end{equation}
where $M_1$ is the average size in the population statistics. Note that to cancel the boundary terms in the integration we need to impose decay conditions on $\psi$. With multiplicative noise $D(x)=D x^2$ for example, we impose 
\begin{align}
\label{eq_cond_decay_psi}
0&= \lim\limits_{x \rightarrow 0} x^2 \psi(x,t) = \lim\limits_{x \rightarrow +\infty} x^2 \psi(x,t)  \\
\label{eq_cond_decay_dpsi}
&=\lim\limits_{x \rightarrow 0} x^3 \partial_{x} \psi(x,t) = \lim\limits_{x \rightarrow +\infty} x^3 \partial_{x} \psi(x,t) \,.
\end{align}
Therefore, if a steady-state is reached, the left hand side of \cref{eq_dt_x_av} is null and 
\be
\label{eq_lam_nu}
\Lambda_t \underset{t \rightarrow \infty}{\rightarrow} \Lambda=\nu \,.
\ee

Second, the lineage-population bias takes a very simple form. 
We show in \cref{app_bias_lin_pop} that for exponential single-cell growth with multiplicative noise, under the decay conditions \cref{eq_cond_decay_psi,eq_cond_decay_dpsi}, the steady-state population distribution $\psi_{\nu}^{b(x)}$ for growth rate $\nu$ and partition kernel $b(x)$ is equal to the steady-state lineage distribution $\phi_{\nu+2D}^{mxb(x)}$ for the modified dynamics $\hat{\nu}=\nu+2D$ and $\hat{b}(x)=mxb(x)$, divided by the size:
\begin{equation}
\label{eq_lin_pop_bias}
\phi_{\nu+2D}^{mxb(x)}(x)=K x \psi_{\nu}^{b(x)}(x) \,,
\end{equation}
where $K=\lp \int_{0}^{\infty} \di x \ x \psi_{\nu}^{b(x)}(x) \rp^{-1}$ is a normalization constant. This result is a particular case of many-to-one formulae for size-structured populations, which also compare the population distribution to the lineage distribution with a biased dynamics \cite{cloez_limit_2017}. In our case, the assumptions of exponential growth and multiplicative noise make the biased dynamics, characterized by $\hat{\nu}$ and $\hat{b}$, explicit and simple.

In the case of deterministic symmetric partitioning, the modified partition kernel  $\hat{b}(x)=mxb(x)=mx\delta(x-1/m)=\delta(x-1/m)=b(x)$ in the single lineage dynamics is equal to the partition kernel $b(x)$ in the population dynamics. Note that this is not true for deterministic asymmetric partitioning. In addition, if there is no noise on the growth, that is if $D=0$, then we recover $\phi(x)=K x \psi(x)$ (\cite{doumic_individual_2021}, p. 41). 

From this last relation, it is straightforward to show that the average snapshot size is larger in lineages than in populations:
\begin{align}
\langle x \rangle_{\rm{lin}} \equiv  N_1 &= M_1+\frac{M_2-M_1^2}{M_1} \nn \\
& \geq M_1 \equiv \langle x \rangle_{\rm{pop}}\,,
\end{align}
where the inequality comes from the positivity of the variance $M_2-M_1^2$ of the size in the population dynamics. 
This result generalizes for any division rate $r$ the inequality observed in \cite{totis_population-based_2021} for particular division rates, provided that there exists a steady-state (if $r=rx^{\alpha}$ then \cref{eq_cond_bet_alp} imposes $\alpha>0$, for more general division rates see conditions on $r$ in \cite{doumic_jauffret_eigenelements_2010}).

A parallel can be drawn between \cref{eq_lin_pop_bias} and the bias derived using the notion of fitness landscape
for a branching tree starting with an ancestor cell of size $x_0$ \cite{genthon_fluctuation_2020}. 
We obtained a lineage-population bias involving a factor $x$, which comes from the correlations between the size $x$ at time $t$ and the number of divisions in the lineage of the cell up to time $t$. Although the frameworks are different, we think that the origin of the factor $x$ in \cref{eq_lin_pop_bias} is also due to these correlations. Moreover, it is clear that if no such correlations are present, the over-representation of lineages with high-reproductive success in a population does not affect the size distribution, so that the lineage and population size distributions are identical.

\section{Exact lineage distributions for deterministic partitioning}
\label{sec_exact_lin}

Exact steady-state solutions to the population \cref{eq_pop_full} have been obtained in the particular case of exponential ($\nu(x)=\nu x$) and noiseless ($D(x)=0$) growth, for deterministic symmetric \cite{hall_functional_1990} and asymmetric \cite{zaidi_asymmetrical_2021} partitioning. 
The same methods can be used to derive the lineage distribution, for which the hypothesis of exponential growth can even be relaxed and replaced by power-law growth rates. 

For deterministic symmetric partitioning between the $m$ daughter cells, we show in \cref{app_dirichlet} that under assumptions \crefrange{eq_pl_r}{eq_cond_bet_alp} and $D(x)=0$, the steady-state solution reads
\begin{align}
\label{eq_series_lin}
\phi(x)&= \frac{C}{x^{\beta}} \sum_{k=0}^{\infty} c_k \exp \lbk - m^{k (\alpha-\beta+1)} \frac{r }{\nu} \ \frac{x^{\alpha-\beta+1}}{\alpha-\beta+1} \rbk \,, 
\end{align}
where the coefficients $c_{k}$ are given in \cref{app_dirichlet} and $C>0$ is a normalizing constant.
For exponential growth, we recover that this lineage distribution is related to the population distribution $\psi(x)$ obtained in (\cite{hall_functional_1990}, eq. 50) by $\phi(x)=x \psi(x)$, which was expected in the light of \cref{sec_exact_bias}. 

It is worth mentioning that this distribution takes a very simple form in the limit of strong control $\alpha \rightarrow +\infty$, where cells divide deterministically when reaching size $1$. %
In this limit, $c_0=1$, $c_1$ tends to $-1$ and all other $c_k$ tend to $0$, such that the lineage size distribution reduces to $\phi(x)=C x^{-\beta}$ for $x \in [1/m,1]$ and $0$ otherwise. This result is analogous to the one for populations: $\psi(x) \propto x^{-2}$ for $x \in [1/2,1]$ and $0$ otherwise, obtained for binary fission and exponential growth \cite{hall_functional_1990,thomas_analysis_2018}. Note also that in this limit, there is no randomness in the dynamics of cell growth and division, and thus the steady-state size distribution is simply the solution to the flux-balance equation $\partial_x [\nu(x) \phi(x,t)]=0$.

In the case of deterministic asymmetric partitioning, for simplicity we choose to focus on binary fission. The volume of the dividing cell is split unequally between the daughters: one inherits a fraction $1/\omega_1$ of the mother size and the other daughter a fraction $1/\omega_2$, with $\omega_1>\omega_2>1$ and $1/\omega_1+1/\omega_2=1$. 
The choice of the protocol to track one of the two daughters is of major importance \cite{jia_cell_2021}. If one chooses to always track the smallest of the two daughters, then the partition kernel is given by: $b(x)=\delta(x-1/\omega_1)$. This is equivalent to the partition kernel for symmetric partitioning between $m$ daughter cells where $m$ is replaced by $\omega_1$, and the size distribution is therefore given by \cref{eq_series_lin}, when replacing $m$ by $\omega_1$. On the other hand, in the random tracking protocol, each cell is tracked at division with probability $1/2$, so that the partition kernel is given by $b(x)=\delta(x-1/\omega_1)/2+\delta(x-1/\omega_2)/2$. In this case, we show in \cref{app_dirichlet_asym} that under assumptions \crefrange{eq_pl_r}{eq_cond_bet_alp} and $D(x)=0$, the size distribution reads
\begin{equation}
\label{eq_series_lin_as}
\phi(x)= \frac{C}{x^{\beta}} \sum_{k=0}^{\infty} \sum_{l=0}^{\infty} c_{k,l} \exp \lbk -  \omega_1^{k (\alpha-\beta+1)}   \omega_2^{l (\alpha-\beta+1)} \frac{r}{\nu} \frac{x^{\alpha-\beta+1}}{\alpha-\beta+1} \rbk \,,
\end{equation}
where the coefficients $c_{k,l}$ are given in \cref{app_dirichlet_asym} and $C>0$ is a normalizing constant.

\subsection{Shapes of the theoretical solutions}

\begin{figure*}
	\centering
	\includegraphics[width=0.4\linewidth]{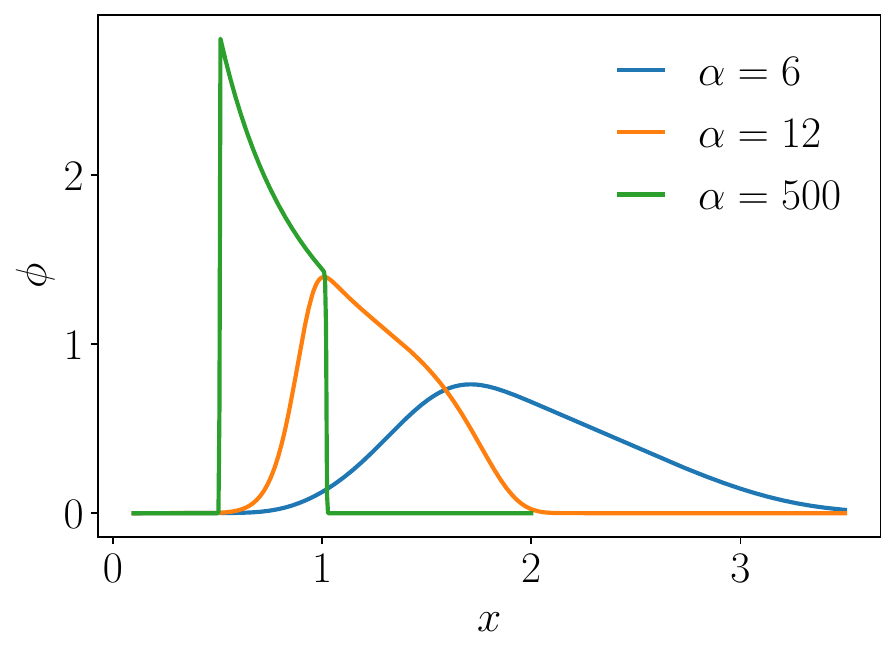}
	\includegraphics[width=0.4\linewidth]{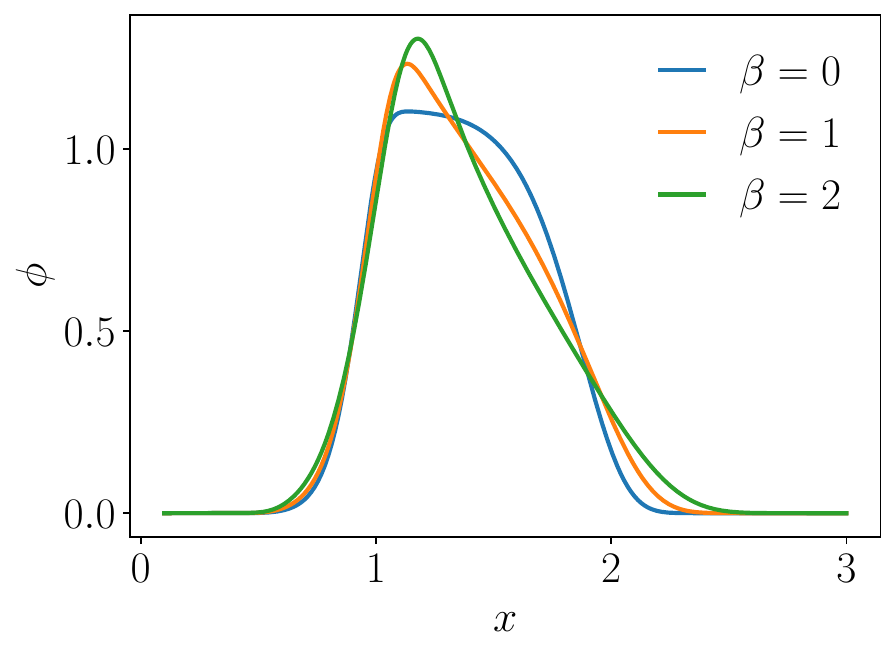}
	\includegraphics[width=0.4\linewidth]{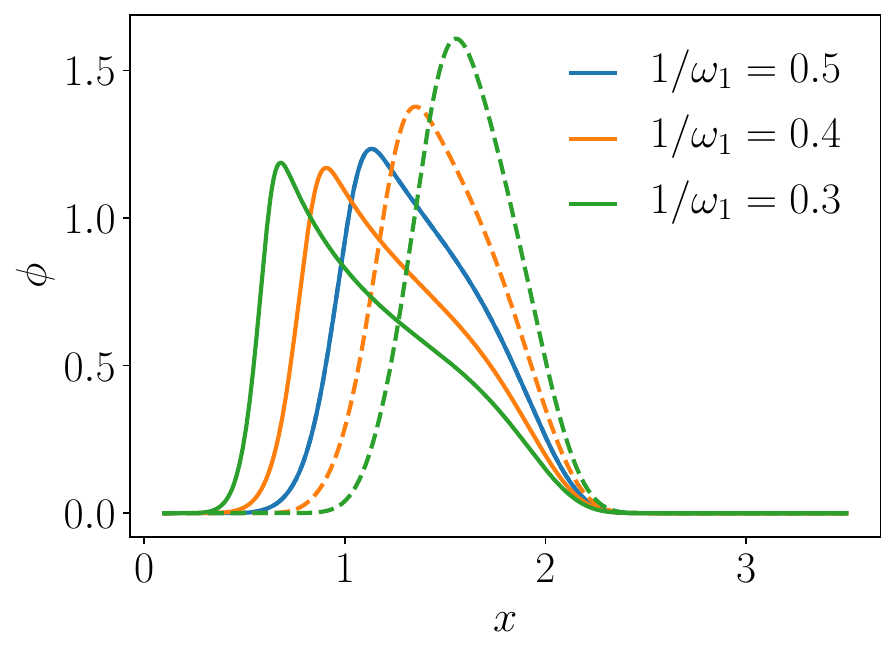}
	\includegraphics[width=0.4\linewidth]{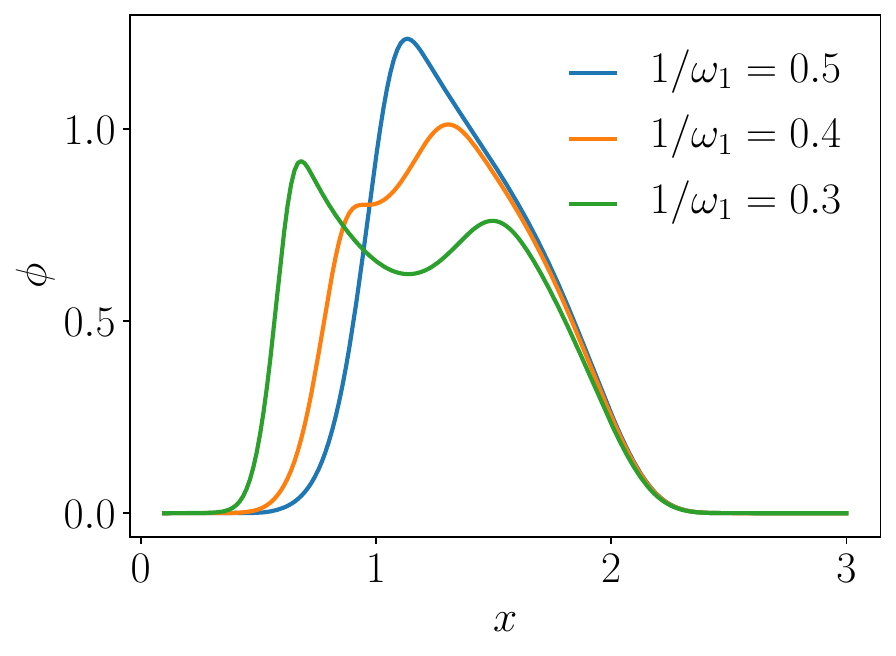}
	\caption{Theoretical lineage distributions for binary fission $m=2$, for deterministic symmetric partitioning on the first row, and deterministic asymmetric partitioning on the second row. 
	On the first row, the distribution is computed with \cref{eq_series_lin}, with $\beta=1$ and $\alpha$ varying on the left, and $\alpha=10$ and $\beta$ varying on the right. 
	For asymmetric partitioning, the left plot shows the distributions obtained with the smallest (\cref{eq_series_lin} with $m=\omega_1$) and largest (\cref{eq_series_lin} with $m=\omega_2=\lp 1- \omega^{-1} \rp^{-1}$) daughter tracking protocols, represented with plain and dotted lines receptively, and the right plot with the random tracking protocol (\cref{eq_series_lin_as}). For both plots, we fixed $\alpha=10$ and $\beta=1$, and varied the asymmetry $\omega_1$. For all four plot we fixed $r/\nu=0.01$. }
	\label{fig_sol_vs}
\end{figure*}

We numerically investigate the influence of the parameters of the model on the analytical steady-state size distributions \cref{eq_series_lin,eq_series_lin_as}, and show the results on \cref{fig_sol_vs}.
The first row corresponds to deterministic symmetric partitioning and the second one to deterministic asymmetric partitioning. On the top left plot, as the strength of the size control $\alpha$ is increased the distribution gets narrower, and in the limit of strong control, division becomes deterministic and $\phi(x)=Cx^{-\beta}$ for $x \in [1/2,1]$. 
On the top right plot, the growth rate power $\beta$ is varied. For $\beta=0$, $\phi$ presents a flat maximum and a fast decline for large size. As $\beta$ increases, the maximum becomes more peaked and the decrease at large size is slowed, which follows from the fact that increasing the growth rate allows cells to reach larger sizes. 
On the second row, $1/\omega_1$ is the volume ratio inherited by the smallest daughter cell, and we show the distributions obtained with the smallest and largest daughter tracking protocols on the left, respectively in plain and dotted lines, and with the random tracking protocol on the right.
On the bottom left plot, the smaller the daughter we follow, the wider the curve on the left hand side. 
Finally, on the bottom right plot, as the asymmetry is increased the curve becomes bimodal, intuitively corresponding to the two subpopulations produced by the smallest and largest daughters at each division.

\subsection{Test on experimental data: parameters inference}

\begin{figure*}
	\includegraphics[width=0.3\linewidth]{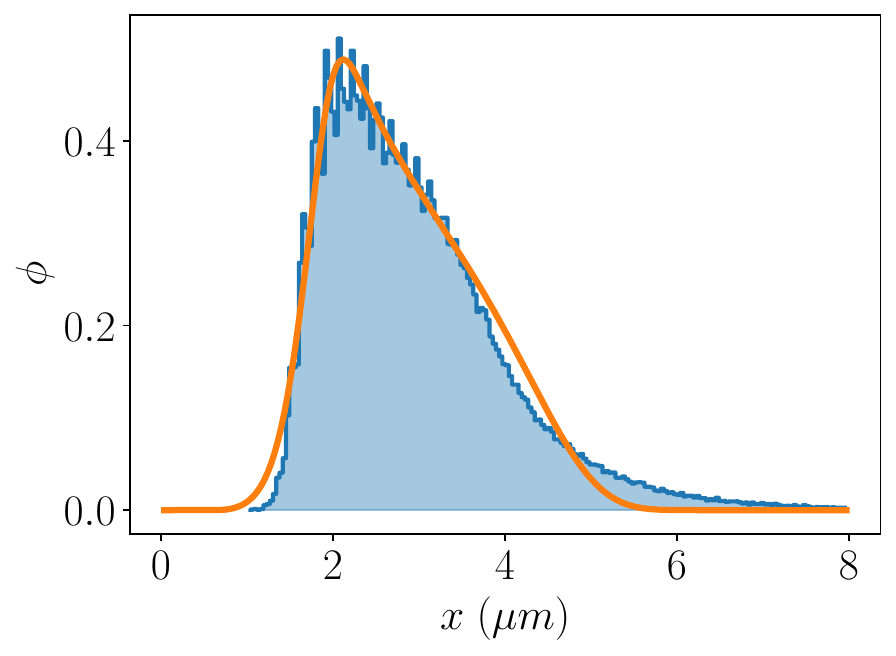}
	\includegraphics[width=0.3\linewidth]{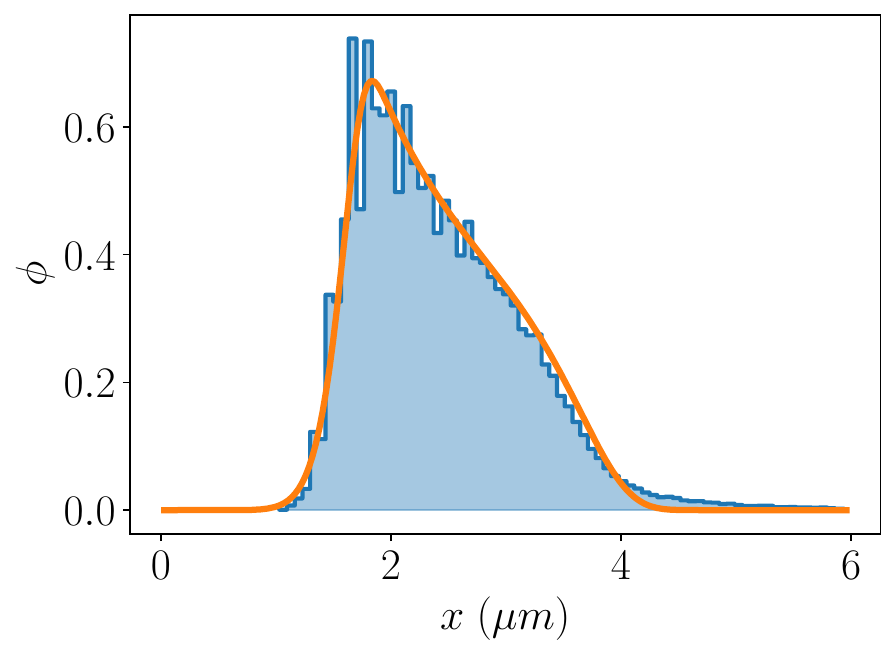}
	\includegraphics[width=0.3\linewidth]{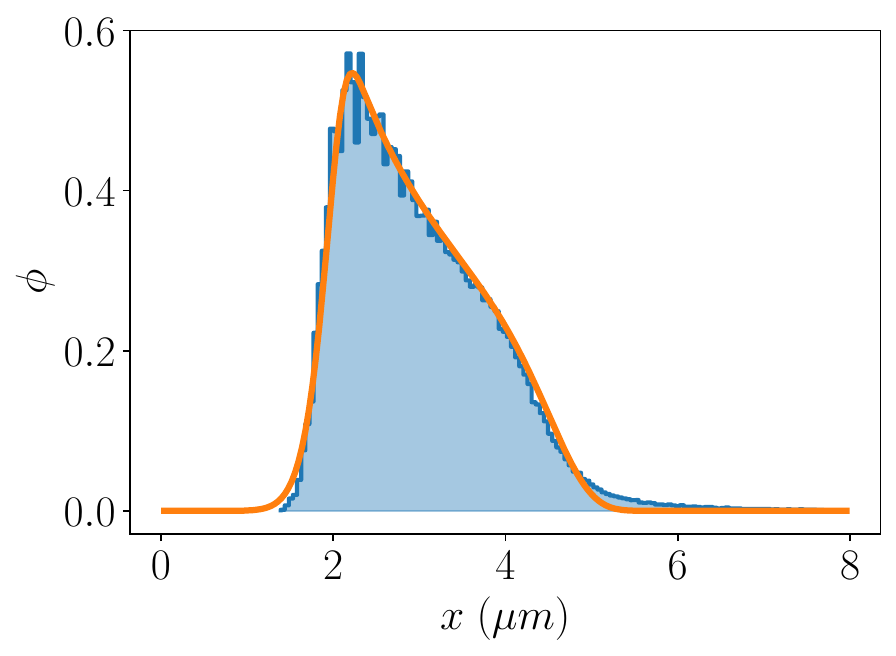}
	\caption{Experimental lineage distributions (blue histograms) for E. coli data from \cite{tanouchi_long-term_2017} in three temperature conditions from left to right: $\SI{25}{\degreeCelsius}$, $\SI{27}{\degreeCelsius}$ and $\SI{37}{\degreeCelsius}$. The best fits (orange curves) are computed with \cref{eq_series_lin} or \cref{eq_series_lin_as}, and the fitting parameters are given in \cref{tab_fit_param}.}
	\label{fig_tan}
\end{figure*}

In experimental systems, the partitioning is stochastic rather than deterministic, however for E. coli data obtained in mother machine \cite{tanouchi_long-term_2017}, the coefficient of variation of the volume ratio at division was found to be smaller than $10\%$ \cite{jia_cell_2021}. This encourages us to test the validity of our theoretical distributions.
We use data from \cite{tanouchi_long-term_2017}, where the size of many independent cell lineages of E. coli has
been recorded every minute over 70 generations at three different temperatures ($\SI{25}{\degreeCelsius}$, $\SI{27}{\degreeCelsius}$, and $\SI{37}{\degreeCelsius}$), precisely 65 lineages for $\SI{25}{\degreeCelsius}$, 54 for $\SI{27}{\degreeCelsius}$, and 160 for $\SI{37}{\degreeCelsius}$. 
We fit the experimental distributions for the three temperatures using the three models: deterministic symmetric partitioning, deterministic asymmetric partitioning with smallest/largest cell tracking and random tracking. For each temperature, the best fit is shown on \cref{fig_tan}, and the fitting parameters are given in \cref{tab_fit_param}. 

\begin{table}
	\begin{tabular}{c|c|c|c}
		& $\SI{25}{\degreeCelsius}$ & $\SI{27}{\degreeCelsius}$ & $\SI{37}{\degreeCelsius}$ \\
		\hline
		\makecell{Tracking \\ protocol} & \makecell{Smallest \\ daughter} & \makecell{Smallest \\ daughter} & \makecell{Smallest \\ daughter} \\
		$\alpha$ & $7.89$ & $11.49$ & $11.92$ \\
		$\beta$ & $1.02$ & $1.26$ & $1.23$ \\
		$1/\omega_1$ & $0.40$ & $0.43$ & $0.43$ \\
		$r/\nu$ & $4.5\times 10^{-5}$ & $3.5 \times 10^{-6}$ & $1.8 \times 10^{-7}$ 	
	\end{tabular}
	\caption{Parameters of the best fits to E. coli data from \cite{tanouchi_long-term_2017} shown on \cref{fig_tan}. In all cases, the best fit was given by the tracking protocol where partitioning is asymmetric and the smallest cell is always followed, given by \cref{eq_series_lin} with $m=\omega_1$. }
	\label{tab_fit_param}	
\end{table}

First of all, we observe that our model is in very good agreement with experiments at $\SI{27}{\degreeCelsius}$ and $\SI{37}{\degreeCelsius}$, and that the fit at $\SI{25}{\degreeCelsius}$ is average but fails to capture the slow decay of the right tail. In particular, our results reproduce the three-stages discussed in \cite{jia_cell_2021}: fast increase for small cells, slow decay for medium-sized cells, and fast decay for large cells.
In the following we analyze the values of the parameters for the condition $\SI{27}{\degreeCelsius}$ and $\SI{37}{\degreeCelsius}$ in particular, given than they provide the best fits to experimental data.
Surprisingly, for all temperatures the best fit is given by deterministic asymmetric partitioning with smaller-daughter tracking protocol, where the daughter cell which is followed inherits a fraction $1/\omega_1=0.43$ of the mother volume. This value is really close to the value $0.44-0.45$ obtained by direct analysis of the sizes at birth and division along the lineages \cite{jia_cell_2021}. 
The strength of the control $\alpha$ tends to increase with temperature, in qualitative agreement with what was found in \cite{jia_cell_2021}, and the ratio $r/\nu$ tends to decrease with temperature. Note that we cannot disentangle the values of $r$ and $\nu$ only from the steady-state profile.
Finally, the power $\beta$ in the growth rate is equal to $1.26$ for $\SI{27}{\degreeCelsius}$ at $1.23$ for $\SI{37}{\degreeCelsius}$, which suggests that in these conditions E. coli grows slightly faster than exponential for large sizes ($x>1$), and slightly slower than exponential for small sizes ($x<1$). This may be linked to the super-exponential growth observed for E. coli in \cite{kar_distinguishing_2021}, where the exponential growth rate $\nu$ increases during the cell cycle.

To conclude, in spite of the stochasticity in partitioning present in experimental systems, our model for deterministic partitioning gives a very good description of the data for two temperature conditions, and a correct fit for the last temperature. It captures the complexity of the distributions, and the inferred parameters show the same trends as the ones obtained from the model proposed by Jia et al. \cite{jia_cell_2021}, based on a $N$-step description of the cell cycle for exponential single-cell growth.
In contrast to this work, (i) our approach allows growth laws that are more general than exponential, and the fit of data from \cite{tanouchi_long-term_2017} revealed slightly non-exponential single-cell growth, which is not accessible in \cite{jia_cell_2021}, (ii) the dependency of the distributions \cref{eq_series_lin} and \cref{eq_series_lin_as} on size is more explicit in our model, and (iii) our model is simpler in that it involves only one step in the cell cycle. 
Even though our result produces a fit for the condition $\SI{25}{\degreeCelsius}$ that is less precise than the one obtained with the $N$-step model, it performs much better than the $N$-step model when $N$ is fixed to $1$, suggesting that models with $N$ steps may not be necessarily minimal.

\section{Asymptotic behavior for general partitioning kernel}
\label{sec_gen_kernel}

In this section, we seek large-size and small-size asymptotic solutions to \cref{eq_lin_full} for deterministic single cell growth ($D(x)=0$) and general partitioning kernels $b$. 
We shall see that the tails of the lineage distribution only depend on the behavior of the division rate, growth rate and partitioning kernel at large and small sizes, like what happens for the population distribution \cite{balague_fine_2013}. Therefore, the results of this section apply to cells obeying more complex growth laws than in the previous section. 

For example, fission yeasts have been observed to follow piece-wise growing patterns \cite{horvath_cell_2013,pesti_cell_2021}, either bi-linear or bi-exponential during the elongation phase. 
Also, the bacterium Corynebacterium glutamicum exhibits asymptotically linear growth \cite{messelink_single-cell_2021}, unlike most bacteria.
In these examples, the growth phases are dictated by the age of the cell, but in size-controlled populations, large (resp. small) cells are on average old (resp. young) cells so that the growth rate at large (resp. small) age is also the growth rate at large (resp. small) size. 
Moreover, E. coli has also be shown to deviate from exponential growth for large and small sizes \cite{robert_division_2014}. 

The power-law assumptions \cref{eq_pl_r,eq_pl_nu} used in the previous section are thus replaced by:
\begin{align}
\label{eq_pl_r_inf}
r(x) &\underset{x \rightarrow 0}{\sim} r_0 x^{\alpha_0} \\
r(x)&\underset{x \rightarrow \infty}{\sim} r_{\infty} x^{\alpha_{\infty}}\\
\nu(x) &\underset{x \rightarrow 0}{\sim} \nu_0 x^{\beta_0}\\
\nu(x)&\underset{x \rightarrow \infty}{\sim} \nu_{\infty} x^{\beta_{\infty}}
\end{align}
with $\alpha_0$, $\alpha_{\infty}$, $\beta_0$, $\beta_{\infty} \geq 0$, and the stability condition \cref{eq_cond_bet_alp} is replaced by \cite{balague_fine_2013}: 
\begin{align}
\label{eq_cond_bet_alp_0}
\alpha_0-\beta_0+1 &> 0 \\
\alpha_{\infty} - \beta_{\infty} +1 &>0  \,.
\end{align}
We describe the partitioning kernel by the following power laws:
\begin{align}
\label{eq_pl_b_0}
b(x)&\underset{x \rightarrow 0}{\sim} b_0 x^{\gamma_0}\\
\label{eq_pl_b_1}
b(x)&\underset{x \rightarrow 1}{\sim} b_1 (1-x)^{\gamma_1} 
\end{align}
with $\gamma_0$, $\gamma_1 >0$, which account for a broad class of kernels defined on $[0,1]$, including the Beta distribution commonly used for volume partitioning \cite{jia_cell_2021}.
Finally, because births of cells inheriting arbitrarily small ratios of the volume of their mothers are allowed by kernel $b$, we need an additional balance condition between growth and loss of volume at division to avoid the formation of cells of vanishing sizes \cite{balague_fine_2013}: 
\be
\label{eq_cond_bet_gam_0}
\gamma_0 - \beta_0+2 >0 \,.
\ee

\subsection{Large size limit}

The large-size population distribution for general growth rate, division rate and partition kernel under assumptions \crefrange{eq_pl_r_inf}{eq_cond_bet_gam_0} has been derived by Balagué et al. \cite{balague_fine_2013}:
\begin{align}
\label{eq_pop_large_non_exp}
\psi(x) &\underset{x \rightarrow \infty}{\sim} \nu(x)^{-1}  \exp \lbk - \int_{}^{x} \di y \ \frac{\Lambda+r(y)}{\nu(y)} \rbk \\
\label{eq_pop_large}
&\underset{x \rightarrow \infty}{\sim} x^{-\beta_{\infty}}  \exp \lbk - \frac{r_{\infty}}{\nu_{\infty}} \frac{x^{\alpha_{\infty}-\beta_{\infty}+1}}{\alpha_{\infty}-\beta_{\infty} +1} - \frac{\Lambda}{\nu_{\infty}} \int_{}^{x} y^{-\beta_{\infty}} \di y  \rbk \,.
\end{align}
This result can be understood intuitively as follows \cite{friedlander_cellular_2008}: if the distribution is decreasing fast enough in the large-size limit, we can neglect the integral term corresponding to the divisions of larger cells in \cref{eq_pop_full}, then the resulting equation is exactly solvable and the solution is \cref{eq_pop_large_non_exp}.

We prove in \cref{app_large_N} and \cref{app_moments_large} that the lineage distribution reads 
\begin{equation}
\label{eq_lin_large}
\phi(x) \underset{x \rightarrow \infty}{\sim} x^{-\beta_{\infty}} \exp \lbk - \frac{r_{\infty}}{\nu_{\infty}} \ \frac{ x^{\alpha_{\infty}-\beta_{\infty}+1}}{ \alpha_{\infty}-\beta_{\infty} +1} \rbk \,,
\end{equation}
which is the population distribution \cref{eq_pop_large_non_exp} when setting $\Lambda=0$. 
The behavior for large sizes is independent of the partition kernel $b$, and it coincides with the large-size behavior for deterministic symmetric partitioning obtained by keeping only the first term the series of \cref{eq_series_lin}. 
 
\begin{figure*}
	\includegraphics[width=0.48\linewidth]{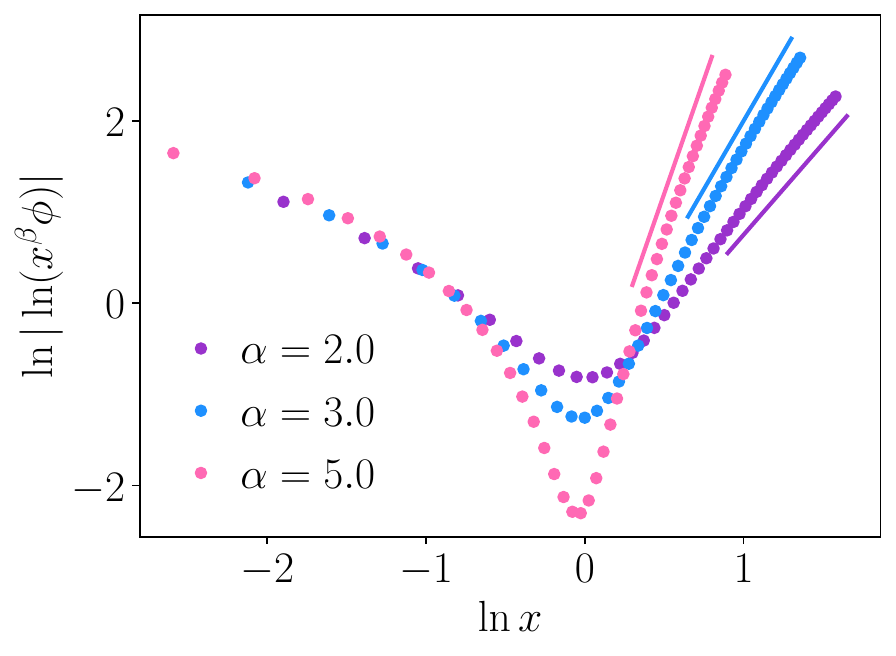}
	\includegraphics[width=0.48\linewidth]{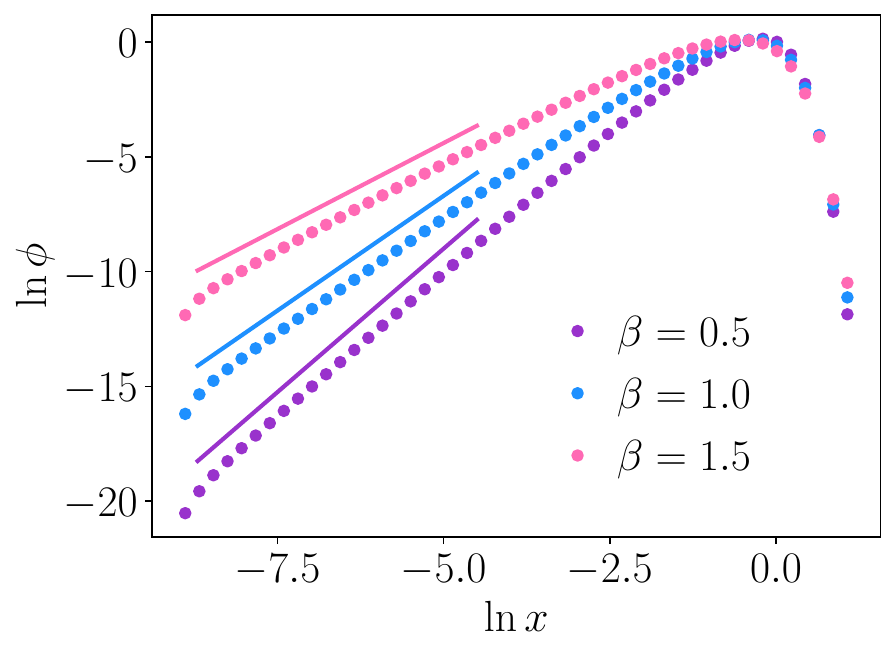}
	\caption{Large and small sizes asymptotic behaviors of the lineage distribution $\phi$. For both plots, we fixed $r(x)=x^{\alpha}$, $\nu(x)=x^{\beta}$, and $b(x)=x^2(1-x)^2/B(3,3)$ (thus $\gamma_0=2$) for all $x$, where $B(x,y)$ is the Beta function. Left: large size limit given by \cref{eq_lin_large} for $\beta=1$, and for three different values of the strength control $\alpha$. The slopes $\alpha-\beta+1=\alpha$ of the solid lines are, from left to right: $5$, $3$ and $2$. 
		Right: small size limit given by \cref{eq_phi_small} for $\alpha=5$ and for three different values of $\beta$. The slopes $\gamma_0+1-\beta=3-\beta$ of the solid lines are, from top to bottom: $1.5$, $2$ and $2.5$. }
	\label{fig_lin_pop_large_x}
\end{figure*}

In order to test this expression, we numerically solve the PBE using a finite difference method with an implicit scheme. Results are shown on \cref{fig_lin_pop_large_x} left for three different values of the strength of size control $\alpha=2$, $3$ and $5$. In all three cases, the large-size behavior is in very good agreement with the theory. 

Comparing \cref{eq_lin_large,eq_pop_large} leads to the following lineage-population biases:
\begin{equation}
\label{eq_bias_large}
\psi(x) \underset{x \rightarrow \infty}{\sim} \begin{cases}
\phi(x) x^{-\Lambda/\nu_{\infty}}& \text{if }\beta_{\infty}=1 \\
\phi(x) \exp \lbk -\frac{\Lambda}{\nu_{\infty}} \frac{x^{1-\beta_{\infty}}}{1- \beta_{\infty}} \rbk & \text{if }\beta_{\infty}\neq 1 \,.
\end{cases}
\end{equation}

If cells grow exponentially with rate $\nu \equiv \nu_{\infty}$ for all sizes and not only for large-sizes, then we showed in \cref{sec_exact_bias} that the population growth rate matches the single cell growth rate $\Lambda=\nu$, so that the lineage-population bias for an arbitrary kernel in the large-size limit is the same as the bias for deterministic symmetric partitioning derived in \cref{sec_exact_bias}. This is not surprising since in this limit the behavior does not depend on the kernel.

For any value $\beta_{\infty} \neq 1$, the exponential function is decreasing with $x$, showing that large cells are under-represented in the population statistics as compared to the lineage statistics, similarly to what happens for exponential growth.

\subsection{Small size limit}

The small-size population distribution for general growth rate, division rate and partitioning kernel under assumptions \crefrange{eq_pl_r_inf}{eq_cond_bet_gam_0} except \cref{eq_pl_b_1} (no assumption on the behavior of the partition kernel $b$ near $x=1$) has been derived by Balagué et al. \cite{balague_fine_2013}:
\begin{equation}
\label{eq_psi_small}
\psi(x) \underset{x \rightarrow 0}{\sim} \begin{cases}
	x^{\gamma_0+1-\beta_0} & \text{if }\beta_0<1 \\
	x^{\gamma_0} & \text{if }\beta_0 \geq 1 \,.
\end{cases}
\end{equation}

We give in \cref{app_small_N} an intuitive argument to understand the piecewise behavior of the population distribution, and show using the same method that, regardless of the value of $\beta_0$, the lineage distribution is given by:
\begin{equation}
\label{eq_phi_small}
\phi(x) \underset{x \rightarrow 0}{\sim} x^{\gamma_0+1-\beta_0} \,.
\end{equation}
This analytical prediction is in perfect agreement with numerical resolutions of the PBE using a finite difference method with an implicit scheme, shown on \cref{fig_lin_pop_large_x} right, for three different values of $\beta=0.5$, $1$ and $1.5$. 

Finally, we obtain the lineage-population bias by comparing \cref{eq_psi_small,eq_phi_small}:
\begin{equation}
\label{eq_bias_small}
\psi(x) \underset{x \rightarrow 0}{\sim} \begin{cases}
	\phi(x) & \text{if }\beta_0<1 \\
	x^{\beta_0-1} \phi(x) & \text{if }\beta_0 \geq 1 \,.
\end{cases}
\end{equation}
When $\beta_0=1$, there is no lineage-population bias as predicted in \cref{sec_exact_bias}. Indeed, $ \psi^{b(x)}(x)$ is equal to $\phi^{2xb(x)}(x)/x \underset{x \rightarrow 0}{\sim} x^{\gamma_0+1}/x = \phi^{b(x)}(x)$, where the factor $x$ and the modified kernel $2xb(x)$ that increases coefficient $\gamma_0$ by $1$ exactly compensate. 
Interestingly, there is no bias for any value $\beta_0 < 1$, which, following the discussion of \cref{sec_exact_bias}, implies that there is no correlation between the size of a cell and the number of divisions along its lineage. Indeed, with deterministic partitioning, the daughter cell inherits half the volume of its mother, so that the only way to reach vanishing sizes is to divide a lot, whereas when all fractions of volume are allowed at division, a cell can also reach small sizes with few divisions if it inherits a small fraction of its mother volume. As a consequence, the correspondence between final size and number of divisions is blurred by the presence of noise in the volume partitioning. 

On the other hand, when $\beta_0 \geq 1$, that is for cells growing slower than exponential in the region of small sizes, the lineage-population bias depends on $\beta_0$. 
Surprisingly, the lineage statistics is biased towards small cells as compared to the population statistics, unlike what we expected from the knowledge of deterministic partitioning. 
This suggests a positive correlation between small sizes and small numbers of divisions, that may be explained as follows. Cells born with very small sizes, because of extremely-asymmetric partitioning allowed by stochastic kernel $b$, grow very slowly because $\beta_0 \geq 1$ and take a long time before reaching average sizes at which they are likely to divide again (which is not the case for $\beta_0<1$). Therefore, these small cells end up with less divisions than average.

\subsection{Validity for the adder model}

Until now, we focused on the \textit{sizer} model, where the division rate is only a function of the size $x$ of the cell. Other models of cell size control have been proposed in the literature, such as the \textit{adder} model which accounts for a broad range of experimental data \cite{taheri-araghi_cell-size_2015,jun_fundamental_2018}. In this model, the division rate depends on both the size $x$ of the cell and the added volume since birth $x-x_b$, with $x_b$ the size at birth, in the following way: $r(x,x_b)=\nu(x) \chi(x-x_b)$. This particular choice of division rate ensures that the distribution of volume added between birth and division is independent of the birth volume. In this section, we show that the asymptotic results derived above remain valid for the adder model. 

To do so, let us first write explicitly a PBE for the adder:
\begin{align}
\label{eq_FP_adder}
\partial_t \psi(x,x_b,t) &= - \partial_x  \lbk \nu(x) \psi(x,x_b,t) \rbk - \lbk \Lambda_t + \nu(x)\chi(x-x_b) \rbk \psi(x, x_b,t) \quad \quad {\rm{for}} \ x>x_b \\
\label{eq_FP_adder_CI}
\nu(x_b) \psi(x_b,x_b,t) &= m \int \frac{\di x'}{x'} \di x_b' \ b(x_b/x') \nu(x')\chi(x'-x_b') \psi(x',x_b',t) \,,
\end{align}
where $\psi(x,x_b,t)$ is the fraction of cells of size $x$ at time $t$ which were born at size $x_b$. 
The lineage equation is obtained from this equation by setting $m=1$ and $\Lambda_t=0$ again. 
Note already a fundamental difference between this equation and \cref{eq_pop_full}: the term accounting for the birth of new cells enters a boundary condition for the adder, because the added volume is reset to $0$ at division. 

For the large-size limit, a direct integration of the steady-state version of \cref{eq_FP_adder} gives
\begin{equation}
\label{eq_pop_large_adder}
\psi(x,x_b) = \psi(x_b,x_b) \frac{\nu(x_b)}{\nu(x)} \exp \lbk - \int_{x_b}^{x} \di y \  \frac{\Lambda}{\nu(y)} + \int_{0}^{x-x_b} \di y \ \chi(y) \rbk \,.
\end{equation}
For any given $x_b$, in the limit where $x \rightarrow + \infty$, we have $x-x_b \sim x$, so that the exponential does not depend on $x_b$ anymore. Thus, the marginal size distribution obeys:
\begin{equation}
\psi(x) \underset{x \rightarrow +\infty}{\sim} \nu(x)^{-1}  \exp \lbk - \int_{}^{x} \di y \ \lp \frac{\Lambda}{\nu(y)} + \chi(y) \rp \rbk \,,
\end{equation}
Finally, this large-size behavior is the same as \cref{eq_pop_large_non_exp} for the sizer, with $r(x)=\nu(x)\chi(x)$. As a consequence, the lineage-population biases in the large-size limit \cref{eq_bias_large} remain valid for the adder model.

In the small-size limit, we saw before that the behavior was controlled by the shapes of the division kernel and the growth rate, which are common in the sizer and adder models, so we anticipate that the results are unchanged. To prove it, we still consider $\nu(x) \underset{x \rightarrow 0}{\sim} \nu_0 x^{\beta_0}$ and $b(x) \underset{x \rightarrow 0}{\sim} b_0 x^{\gamma_0}$, and follow the method used for the sizer presented in \cref{app_small_N}. Multiplying \cref{eq_FP_adder} by $x^k$ and integrating over $x$ and $x_b$ leads after simple manipulations to:
\begin{equation}
\label{eq_mom_pop_small_adder}
\nu_0(m L_k -1) \int \di x \di x_b \ x^{\beta_0+k} \chi(x-x_b) \psi(x,x_b) = \Lambda M_k - \nu_0 k M_{k+\beta_0-1} \,,
\end{equation}
where $M_k=\int_{0}^{\infty} \di x \ x^k \psi(x)  \equiv \int_{0}^{\infty} \di x \di x_b \ x^k \psi(x,x_b)$ is the $k$-th moment of the marginal size distribution $\psi(x)$.
Now we suppose that there is a $\delta_0 \geq 0$ such that $\chi(x) = O(x^{\delta_0})$ when $x \rightarrow 0$, meaning that the division rate per unit volume $\chi$ is growing as a power law or slower. In this case, the integral in the left hand side is smaller than $M_{k+\beta_0+\delta_0}$, and thus the integral does not diverge when $M_{k+\beta_0+\delta_0}$ does not diverge. The rest of the proof is the same as for the sizer, where $\beta_0+\delta_0$ plays the role of $\alpha_0$. Finally, the small-size lineage-population biases \cref{eq_bias_small} remain true for the adder model. 

Before closing this section, we would like to draw the attention of the reader on the fact that in none of the different lineage-population biases derived in the previous sections does the division rate appear explicitly. In this section we showed that they hold for the adder model, which is a particular choice of two-variable division rate. This observation suggests that these biases could be correct for a much broader class of division rates, possibly involving other variables.

\section{Noisy single-cell exponential growth}
\label{sec_noise}

\begin{figure*}
	\includegraphics[width=0.8\linewidth]{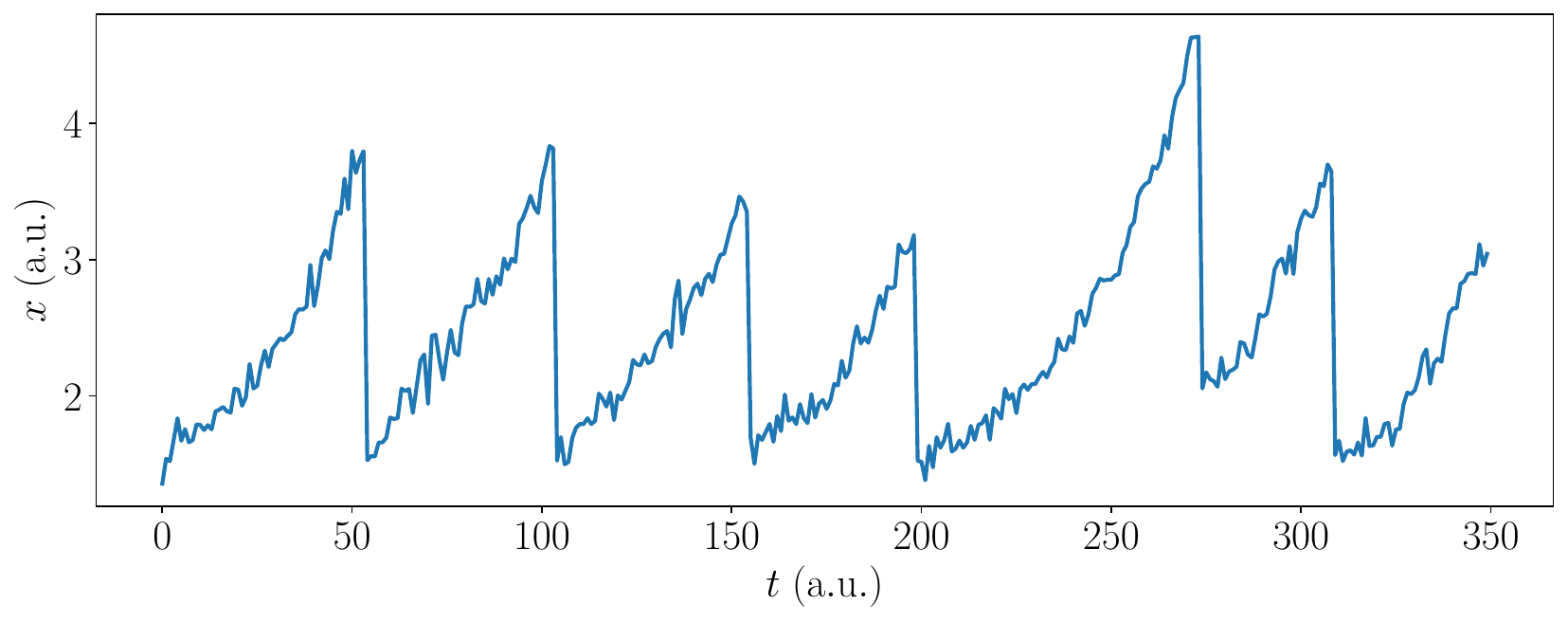}
	\caption{Example of size evolution versus time for a single lineage from \cite{tanouchi_long-term_2017}, in the condition $\SI{27}{\degreeCelsius}$.}
	\label{fig_tan_traj}
\end{figure*}

As discussed in the introduction, bacterial size trajectories show some fluctuations around exponential growth. This is illustrated on \cref{fig_tan_traj}, where we plot a single cell size trajectory using E. coli data from \cite{tanouchi_long-term_2017}. 
To account for these fluctuations, we now introduce in-cycle multiplicative noise $D(x)=D x^2$ on top of single-cell exponential growth $\nu(x)=\nu x$.

To our knowledge, exact solutions to the steady-state version of \cref{eq_pop_full} were obtained for deterministic partitioning and for specific growth rates $\nu(x)$, division rates $r(x)$ and diffusion coefficients $D(x)$ only. 
For instance, it was solved for constant functions $\nu(x)$,  $r(x)$ and $D(x)$, both for deterministic symmetric \cite{efendiev_functional_2018} and asymmetric \cite{efendiev_asymmetric_2018} partitioning; and for deterministic asymmetric partitioning, exponential growth $\nu(x)=\nu x$, multiplicative noise $D(x)=Dx^2$ and quadratic division rate $r(x)=r x^2$ \cite{zaidi_probability_2016}. In this last case, the solution is a series of modified Bessel functions, generalizing the Dirichlet series obtained when there is no diffusion, and arising from the quadratic division rate which turns \cref{eq_pop_full} into a modified Bessel equation. Therefore, it seems difficult to generalize this method to more general power law division rates. 
In this section we seek asymptotic solutions for large sizes in steady-state (see \cite{marguet_law_2019,bansaye_nonconservative_2022} for convergence to steady-state), for more general divisions rates $r(x)\underset{x \rightarrow \infty}{\sim} r_{\infty} x^{\alpha_{\infty}}$ and partition kernels, when imposing decay conditions \cref{eq_cond_decay_psi,eq_cond_decay_dpsi} so that $\Lambda=\nu$ (\cref{eq_lam_nu}).

We show in \cref{app_large_N_noise} and \cref{app_moments_large} that the lineage and population steady state distributions are equivalent in the large size limit and given by

\begin{equation}
\label{eq_noise_phi}
\psi(x) \underset{x \rightarrow \infty}{\sim} \phi(x) \underset{x \rightarrow \infty}{\sim} x^{\frac{\nu}{2 D}-\frac{3}{2}-\frac{\alpha_{\infty}}{4}} \exp \lbk - \frac{2 }{\alpha_{\infty}} \sqrt{\frac{r_{\infty}}{D}}  \ x^{\frac{\alpha_{\infty}}{2}} \rbk \,.
\end{equation}

Just like the case of deterministic growth, the large-size behaviors of $\phi$ and $\psi$ are independent of the partitioning kernel.
In the case $\alpha_{\infty}=2$, we recover the result obtained in \cite{zaidi_probability_2016} (up to a missing factor $\sqrt{x}$ due to a typo).

\begin{figure}
	\includegraphics[width=0.5\linewidth]{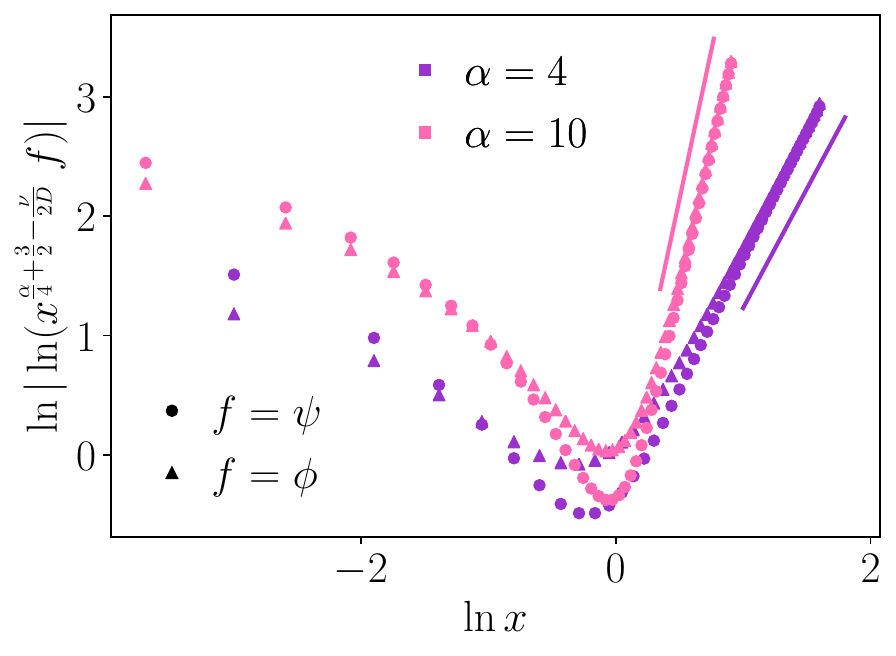}
	\caption{Large size asymptotic behaviors of the lineage and population distributions $\phi$ and $\psi$ given by \cref{eq_noise_phi}. We fixed $r(x)=x^{\alpha}$, $\nu(x)=x$, $D=0.4$ and $b(x)=x^2(1-x)^2/B(3,3)$ for all $x$, and show two different values of the strength control $\alpha$. The slopes $\alpha/2$ of the solid lines are, from left to right: $5$ and $2$. }
	\label{fig_large_x_noise}
\end{figure}

We numerically solve the PBE with the diffusive term using a finite difference method, both for the lineage statistics and the population statistics. The scheme is implicit in the first case and hybrid in the second: all terms are implicit except $\Lambda_t$, given by \cref{eq_def_lam}, which is explicitly computed. Results are shown on \cref{fig_large_x_noise}, for two different values $\alpha=4$ and $\alpha=10$ of the size control strength. In both cases, the population and lineage distributions coincide in the large-size limit and align with the theoretical prediction \cref{eq_noise_phi}.

Unlike the case of deterministic growth discussed in \cref{sec_gen_kernel}, no lineage-population bias is observed here. 
This is coherent with \cref{eq_lin_pop_bias}, where the bias towards smaller cells accounted for by the factor $x$ and the effective growth rate $\nu+2D$ exactly compensate. Indeed, one easily check from \cref{eq_noise_phi} that $\phi_{\nu+2D}(x)/x=\phi_{\nu}(x)$.
Similarly to what happens for small sizes in presence of a stochastic kernel, the lineage-population bias is killed by the presence of multiplicative noise. When there is no noise, only cells that divided few times can reach large sizes, which imposes correlations between the number of divisions and the final size. Here however, this correlation is canceled because the number of divisions can be balanced by the noisy growth: large cells can come from lineages with numerous divisions if they grew faster on average than the deterministic growth at rate $\nu$.

\section{Conclusion}

The recent development of mother machine devices revived the interest in the statistical comparison between lineage measurements and population snapshots. The unprecedented amount of single-cell data offers new insights on the way cells regulate their cycle and maintain size homeostasis. It is hence fundamental to quantify the statistics obtained in single-lineage setups and to understand how they differ from classical population snapshots. 

Such lineage-population statistical biases have been studied for example for cell age \cite{powell_growth_1956,thomas_making_2017}, for the number of divisions \cite{nozoe_inferring_2017}, or for cell size at birth \cite{thomas_analysis_2018}. However, despite the importance of cell size in the understanding of the regulation of the cell cycle, and although some works have been done in this direction in simple cases \cite{genthon_fluctuation_2020,totis_population-based_2021}, quantitative lineage-population biases for size statistics are lacking in general contexts. 
In this article, we addressed this question and the impact of the different sources of stochasticity on this bias. 

For exponential single-cell growth, we showed that for any division rate and partitioning kernel the population distribution is proportional to the lineage distribution with a modified dynamics, divided by the size $x$. 
This bias is reminiscent of the correlations between the size and the number of divisions \cite{genthon_fluctuation_2020}, and implies that for deterministic symmetric division, cells are on average smaller in population than in lineage, which generalizes results from \cite{totis_population-based_2021} obtained for particular division rates. 

In a recent study, Jia et. al \cite{jia_cell_2021} derived an analytical lineage distribution for size-controlled exponentially-growing cells with a model of cell-cycle split in $N$ stages, and successfully compared it with experimental data to infer the model parameters. 
On our side, we obtained analytical lineage distributions for deterministic partitioning both symmetric and asymmetric, for general power-law single-cell growth rates, which include exponential growth, that are complementary to the solution from \cite{jia_cell_2021}. We compared our solution to the same set of experimental mother-machine data, and found a good agreement despite the slight stochasticity of the partitioning in these data. The relevant parameters of the model, such as the strength of the size control and the asymmetry of the division can be obtained from the fit, and are in agreement with those obtained in \cite{jia_cell_2021}. In addition, the fact that our model allows single-cell growth rates more complex than exponential revealed slightly non-exponential growth laws for E. coli, that were not accessible in \cite{jia_cell_2021}.

When relaxing the hypothesis of deterministic partitioning, the small and large-size tails of the population distribution have been previously derived by Balagué et al. in \cite{balague_fine_2013}, and we derived in this article the same tails for the lineage distribution. Note that if, in the large-size limit, the lineage behavior is in the end given by the population behavior when following only one cell (by canceling the population growth rate in the solution), this is not true for the small-size behavior, where a non-trivial difference based on the growth rate of small cells appears, and which can thus not be deduced from the population behavior.  
By comparing the distributions, we obtained the lineage-population bias in the two limits, and showed that they only depend on the single-cell growth rate but are independent of the division rate and the partition kernel. 

An important extension of these results is the validity of the asymptotic behaviors and biases for the adder mechanism, which states that cells division is triggered by the increment of volume since birth, and which is increasingly seen as the most relevant model of cell size control. To our knowledge, size distributions have not been obtained before for the two-variable adder model (size and size at birth) whether in population or lineage. Since the biases do not depend explicitly on the division rate, we expect them to remain true for more general mechanisms of division, even beyond the adder model. 

Since single-growth is not indeed a deterministic process, we introduced diffusivity around exponential growth via multiplicative noise, to account for fluctuations observed in data while ensuring size positivity. This source of stochasticity has been much less studied than the cell-to-cell variability in growth rate. To our knowledge, the only article assessing the impact of multiplicative noise on size distribution is \cite{zaidi_probability_2016}, only for a quadratic division rate. We extended the large-size behavior of the population distribution from this reference to the case of more general power-law division rates, and doing the same for the lineage distribution we derived the lineage-population bias in the presence of noisy growth. 

Finally, this analysis revealed that introducing sources of stochasticity can cancel the lineage-population bias, thus blurring the correlations between the size and the number of divisions undergone by the cell. Indeed, when considering stochastic partitioning kernels and small cells growing faster than exponential, the small-size lineage-population bias is canceled; and when introducing fluctuations around deterministic exponential growth, the lineage-population bias in the large-size limit is canceled.

This work can be extended in several directions. 
First, experimental inference suggests a non-trivial behavior of the division rate at large sizes \cite{robert_division_2014}, even though the estimation in this region could be unreliable because of the lack of statistics. Therefore, it would be useful to relax the hypothesis of a power-law division rate.
Second, we focused on a one-variable model, with the exception of the two-variable adder model, and it would be interesting to investigate more complex models with $n$ variables. For example, modeling cell-to-cell variability in growth imposes to treat the single-cell growth rate $\nu$ as a second random variable. 
Finally, constant-population experiments, such as the dynamics cytometer \cite{hashimoto_noise-driven_2016}, may not all be well described by a uniform dilution rate, but rather by a dilution rate dependent on size, generation, spatial position, ... 
In this case, the size distribution obtained would not be the same as in a freely-growing population but 
would bear the mark of the dilution protocol.

\begin{acknowledgments}
	
	I warmly thank Marie Doumic for her essential help with the mathematical literature, Jérémie Unterberger and David Lacoste for their careful reading of the manuscript and for useful discussions, and the anonymous referees for their valuable comments.
	
\end{acknowledgments}

\appendix

\onecolumngrid

\section{Lineage-population bias for exponentially-growing cells}
\label{app_bias_lin_pop}

In this section, we consider the case of exponential growth $\nu(x)=\nu x$ with multiplicative noise $D(x)=Dx^2$, under the decay conditions \cref{eq_cond_decay_psi,eq_cond_decay_dpsi} so that in steady-state $\Lambda=\nu$, and for general division rate $r$ and partitioning kernel $b$. In this case, the steady-state \cref{eq_pop_full,eq_lin_full} read:
\begin{align}
0 &= -\nu  \partial_x [x \psi(x)] + D \partial_{x^2} [x^2 \psi(x)] - \lbk r(x) + \nu \rbk \psi(x) + m \int \frac{dx'}{x'} b(x/x')
r(x') \psi(x') \\
\label{eq_lin_exp_bias}
0 &= -\nu  \partial_x [x \phi(x)] + D \partial_{x^2} [x^2 \phi(x)] - r(x) \phi(x) + \int \frac{dx'}{x'} b(x/x')
r(x') \phi(x') \,.
\end{align}
We multiply the population equation by $x$, and recast it for the function $q(x)=x\psi(x)$:
\begin{equation}
0 = -\nu x \partial_x q(x) + D x \partial_{x^2} [x q(x)] - \lbk r(x) + \nu \rbk q(x) + \int \frac{dx'}{x'} \frac{mx}{x'}b(x/x')
r(x') q(x') \,.
\end{equation}
We identify the derivative of a product $-\nu x \partial_x q(x) - \nu q(x) = -\nu \partial_x [xq(x)]$, and show straightforwardly that $D x \partial_{x^2} [x q(x)]= D \partial_{x^2} [x^2 q(x)]-2 D \partial_{x} [ x q(x)]$, so that the second term is absorbed in the first-order derivative describing exponential growth:
\begin{equation}
0 = -(\nu+2D) \partial_x [xq(x)] + D \partial_{x^2} [x^2 q(x)] - r(x) q(x) + \int \frac{dx'}{x'} \frac{mx}{x'}b(x/x')
r(x') q(x') \,.
\end{equation}
This equation is \cref{eq_lin_exp_bias}, obeyed by the lineage distribution $\phi$ with modified growth rate $\hat{\nu}=\nu+2D$ and partition kernel  $\hat{b}(x)=mxb(x)$, therefore $q(x)_{\nu}^{b(x)}$ is proportional to $\phi_{\nu+2D}^{mxb(x)}$:
\begin{equation}
\phi_{\nu+2D}^{mxb(x)}(x)=K x \psi_{\nu}^{b(x)}(x) \,,
\end{equation}
with $K=\lp \int_{0}^{\infty} \di x \ x \psi_{\nu}^{b(x)}(x) \rp^{-1}$ a normalization constant.
Importantly, $\hat{b}$ is a proper kernel, which is normalized as a consequence of the conservation of volume of the original kernel $b$: $\int_{0}^{1} \di x \ \hat{b}(x) = m \int_{0}^{1} \di x \ x b(x)=1$.

\section{Exact lineage solution for deterministic partitioning}

In this section, we seek exact steady-state solutions to \cref{eq_lin_full} for deterministic single cell growth ($D=0$), deterministic volume partitioning, both symmetric or asymmetric, and under assumptions \crefrange{eq_pl_r}{eq_cond_bet_alp} for the growth and division rates. 

\subsection{Symmetric partitioning}
\label{app_dirichlet}

We follow the method proposed in \cite{hall_functional_1990} starting from the steady-state equation:
\begin{equation}
\label{app_lin_FP}
[\nu x^{\beta} \phi(x)]'  = -rx^{\alpha} \phi(x) + m^{1+\alpha} r x^{\alpha} \phi(mx) \,,
\end{equation}
and we define $Z(x)=x^{\beta} \phi(x)$, then
\begin{equation}
\label{}
Z'(x)  = \frac{r}{\nu}x^{\alpha-\beta} \lbk - Z(x) + m^{1-\beta+\alpha} Z(mx) \rbk \,.
\end{equation}
We now define $u=r x^{\alpha-\beta+1}/[\nu (\alpha-\beta+1)]$ and $Y(u)=Z(x)$, so that the equation on $Y$ reads:
\begin{equation}
\label{}
Y'(u)+Y(u) = m^{1-\beta+\alpha} Y(m^{1-\beta+\alpha}u) \,.
\end{equation}
Since $1-\beta+\alpha>0$, the solution to this equation is \cite{hall_functional_1989}:
\begin{align}
\label{app_series_lin_Y}
Y(u)&= C\sum_{k=0}^{\infty} c_k \exp \lbk -  m^{k (\alpha-\beta+1)}  u \rbk \\
c_0&=1 \\
c_k&=\frac{(-1)^k m^{k (\alpha-\beta+1)}}{ \prod_{j=1}^{k}(m^ {j(\alpha-\beta+1)}-1)} & k \geq 1 \,,
\end{align}
where $C>0$ is a normalizing constant.

Reverting to the notation in $x$ with the function $\phi$ gives
\begin{equation}
\phi(x)= \frac{C}{x^{\beta}} \sum_{k=0}^{\infty} c_k \exp \lbk - m^{k (\alpha-\beta+1)} \frac{r }{\nu} \ \frac{x^{\alpha-\beta+1}}{\alpha-\beta+1} \rbk \,.
\end{equation}

\subsection{Asymmetric partitioning}
\label{app_dirichlet_asym}

For simplicity we consider binary fission ($m=2$), with asymmetric partitioning: $b(x)=\delta(x-1/\omega_1)/2+\delta(x-1/\omega_2)/2$, where $\omega_1 > \omega_2 >1$ and $1/\omega_1+1/\omega_2=1$. Starting from:
\begin{equation}
\label{app_lin_FP_asym}
[\nu x^{\beta} \phi(x)]'  = -rx^{\alpha} \phi(x) + \frac{rx^{\alpha}}{2} \lbk \omega_1^{1+\alpha} \phi(\omega_1 x) + \omega_2^{1+\alpha} \phi(\omega_2 x)\rbk \,,
\end{equation}
and making the same two changes of variables as for the symmetrical case, the equation reads
\begin{equation}
\label{}
Y'(u)+Y(u) = \frac{1}{2} \lbk \Omega_1 Y(\Omega_1 u) + \Omega_2 Y(\Omega_2 u) \rbk \,,
\end{equation}
where we defined $\Omega_i=\omega_i^{1-\beta+\alpha}$ for $i\in \{1,2\}$. 
Since $1-\beta+\alpha>0$, the solution to this equation is \cite{suebcharoen_asymmetric_2011}:
\begin{align}
Y(u)&= C \sum_{k=0}^{\infty} \sum_{l=0}^{\infty} c_{k,l} \exp \lbk -  \Omega_1^k   \Omega_2^l u \rbk \\
c_{0,0}&=1  \\
c_{k,0}&=\frac{(-1)^k\Omega_1^k}{2^k \prod_{j=1}^{k}(\Omega_1^j-1)} \\
c_{0,l}&=\frac{(-1)^l\Omega_2^l}{2^l \prod_{j=1}^{l}(\Omega_2^j-1)} \\
c_{k,l}&=\frac{\Omega_1 c_{k-1,l} + \Omega_2 c_{k,l-1}}{2-2\Omega_1^k \Omega_2^l} \,,
\end{align}
where $C>0$ is a normalizing constant.

Reverting to the original notations gives
\begin{equation}
\phi(x)= \frac{C}{x^{\beta}} \sum_{k=0}^{\infty} \sum_{l=0}^{\infty} c_{k,l} \exp \lbk -  \omega_1^{k (\alpha-\beta+1)}   \omega_2^{l (\alpha-\beta+1)} \frac{r}{\nu} \frac{x^{\alpha-\beta+1}}{\alpha-\beta+1} \rbk \,.
\end{equation}

\section{Asymptotic distributions for stochastic partitioning}

In this section, we seek the steady-state asymptotic behaviors of the size distributions in the large and small size limits for general partitioning kernels. Let us first make two general comments.
First, when the limit $k \rightarrow  +\infty$ (resp. $k \rightarrow  -\infty$) is considered, corresponding to the large size (resp. small size) behavior, the integrals of the type $\int_{0}^{\infty} \di x x^k f(x)$ are dominated by the behavior of the function $f$ as $x \rightarrow +\infty$ (resp. $x \rightarrow 0$). Therefore, when for example $r(x) \underset{x \rightarrow +\infty}{\sim} rx^{\alpha}$, we write $\int_{0}^{\infty} \di x x^k r(x) \phi(x) \underset{k \rightarrow +\infty}{\sim} \int_{0}^{\infty} \di x x^k r x^{\alpha} \phi(x) = r N_{\alpha+k}$.
Second, the following transformation is used to isolate the moments $L$ of the kernel $b$ \cite{cheng_scaling_1988}: 
\begin{align}
&\int_{0}^{\infty} \di x \ x^k \int_{x}^{\infty} \di y \ b(x/y) y^{\alpha-1}  \phi(y) \nn\\
&= \int_{0}^{\infty} \di y \ y^{\alpha-1}  \phi(y)\int_{0}^{y} \di x \ x^k b(x/y)  \nn\\
&= \int_{0}^{\infty} \di y \ y^{k+\alpha} \phi(y) \int_{0}^{1} \di u \ u^k b(u) \nn\\
\label{eq_approx_L}
&= N_{\alpha+k} \ L_k \,,
\end{align}
where we went from the second to the third line with the change of variable $u=x/y$.

\subsection{Large size limit}

In the following subsections, since only the large-size limit is investigated, for better readability we drop the subscript $\infty$ for $r$, $\nu$, $\alpha$ and $\beta$, and the limit $\to \infty$ is always understood as $\to + \infty$

\subsubsection{Deterministic growth}
\label{app_large_N}

We first consider the case of deterministic growth ($D=0$) with assumptions \crefrange{eq_pl_r_inf}{eq_cond_bet_gam_0} on the different rates, and follow the method proposed in \cite{cheng_scaling_1988} for fragmentation processes. We multiply the steady-state \cref{eq_lin_full} by $x^{k-\beta+1}$ and integrate over $x$, to recast the PBE as a recursion relation on the moments of the distribution:
\begin{equation}
\label{eq_rec_mom_lin_deter}
N_{k+\alpha-\beta+1} \underset{k \rightarrow \infty}{\sim} \frac{\nu}{r} \ \frac{k- \beta +1}{1 - L_{k-\beta +1}} N_k \,.
\end{equation}
For simplicity we define $\rho=\alpha-\beta+1 > 0$, and $n$ such that $k=n\rho$, leading to 
\begin{equation}
N_{(n+1) \rho} \underset{n \rightarrow \infty}{\sim} \frac{\nu}{r} \ \frac{ n \rho- \beta +1 }{1 - L_{n \rho-\beta +1}} N_{n \rho} \,.
\end{equation}
Iterating this relation leads to the general term:
\begin{equation}
\label{eq_general_N_large}
N_{n \rho} \underset{n \rightarrow \infty}{\sim} N_{\rho} \lp \frac{\nu}{r} \rp^{n-1} \prod_{j=1}^{n-1} \ \frac{ j \rho- \beta +1 }{1 - L_{j \rho-\beta +1}} \,.
\end{equation}
We compute the numerator as 
\begin{align}
\prod_{j=1}^{n-1} \lp j \rho- \beta +1 \rp &= \rho^{n-1} (n-1)! \prod_{j=1}^{n-1} \lp 1- \frac{\beta -1}{j \rho } \rp \\
& \underset{n \rightarrow \infty}{\sim} \rho^{n-1} (n-1)! (n-1)^{\frac{1-\beta}{ \rho }} \,,
\end{align}
where we used that $\prod_{j=1}^{n} \lp 1- \frac{a}{j} \rp \sim n^{-a}$ as $n \rightarrow \infty$.

We now show that the moments $L_{j \rho-\beta +1}$ of the partition kernel can be neglected in this limit:
\begin{align}
L_k & \underset{k \rightarrow \infty}{\sim} b_1 \int_{0}^{1} \di x \ x^k  (1-x)^{\gamma_1} \\
&\underset{k \rightarrow \infty}{\sim} k^{-(\gamma_1+1)} \,,
\end{align}
where we recognize the Beta function $B(k+1,\gamma_1+1)=\int_{0}^{1} \di y \ y^k (1-y)^{\gamma_1}$, whose asymptotic behavior when only one of the two parameters tends to infinity (here $k$) is given by: $B(k+1,\gamma_1+1)\underset{k \rightarrow \infty}{\sim} \Gamma(\gamma_1+1) k^{-(\gamma_1+1)}$. 
The product in the denominator of \cref{eq_general_N_large} is therefore given by
\begin{align}
\ln \prod_{j=1}^{n-1} (1 - L_{ j \rho- \beta +1}) & = \sum_{j=1}^{n-1} \ln (1 - L_{ j \rho- \beta +1}) \\
& \underset{n \rightarrow \infty}{\sim} - \sum_{j=1}^{n-1} \lp  j \rho- \beta +1 \rp^{-(\gamma_1+1)}\,,
\end{align}
where the second line is obtained by a first-order expansion of the natural logarithm. Finally, since $\gamma_1 > 0$, this series is converging when $n \rightarrow \infty$.

The general term then reads
\begin{equation}
N_{n \rho} \underset{n \rightarrow \infty}{\sim} \lp\frac{\rho \nu}{r}\rp^{n-1} (n-1)! (n-1)^{\frac{1-\beta}{\rho}} \,.
\end{equation}
We next use Stirling approximation: $n! \underset{n \rightarrow \infty}{\sim} \sqrt{2 \pi n} (n/e)^n$, switch back to $k=n\rho$ and replace $\rho$:
\begin{equation}
\label{eq_large_N}
N_k \underset{k \rightarrow \infty}{\sim} \lp \frac{\nu k}{r e } \rp^\frac{k}{\alpha-\beta +1} k^{\frac{1-\beta}{\alpha-\beta +1}-\frac{1}{2}} \,.
\end{equation}
The inverse Mellin transform of this moment is obtained in \cref{app_moments_large}.

\subsubsection{Stochastic growth}
\label{app_large_N_noise}

We examine the case of exponential growth $\nu(x)=\nu x$ with multiplicative noise $D(x)=Dx^2$, under the decay conditions \cref{eq_cond_decay_psi,eq_cond_decay_dpsi} so that in steady-state $\Lambda=\nu$ (\cref{eq_lam_nu}), for general division rates of the form $r(x)\underset{x \rightarrow \infty}{\sim} r x^{\alpha}$.
Following the same steps as for the deterministic growth, we multiply the steady-state \cref{eq_pop_full,eq_lin_full} by $x^k$ and integrate over $x$:
\begin{align}
M_{k+\alpha} &\underset{k \rightarrow \infty}{\sim} \frac{1}{r} \ \frac{\nu (k-1) + D k (k-1)}{1 - m L_{k}} M_k \\
N_{k+\alpha} &\underset{k \rightarrow \infty}{\sim} \frac{1}{r} \ \frac{\nu k + D k (k-1)}{1 - L_{k}} N_k \,.
\end{align}
As for the deterministic case, the moments $L_{k}$ are negligible for large $k$, so that the moments $M_{k+\alpha} $ and $N_{k+\alpha} $ differ only by their numerators. We conduct the calculations for the lineage distribution first and then show why the difference in the numerators does not affect the general moment.

We define $n$ such that $k=n\alpha$, and iterate the relation to obtain the general term:

\begin{equation}
N_{n \alpha}\underset{n \rightarrow \infty}{\sim} N_{\alpha} \lp \frac{1}{r} \rp^{n-1} \prod_{j=1}^{n-1} \ \frac{\nu j \alpha + D j\alpha (j\alpha-1)}{1 - L_{j \alpha}} \,.
\end{equation}
The numerator is computed as:
\begin{align}
\prod_{j=1}^{n-1} \lbk \nu j \alpha + D j\alpha (j\alpha-1) \rbk &= (\alpha^2 D)^{n-1} (n-1)!^2 \prod_{j=1}^{n-1} \lp 1- \frac{D -\nu}{j \alpha D} \rp \\
\label{eq_large_N_num_noise}
& \underset{n \rightarrow \infty}{\sim} (\alpha^2 D)^{n-1} (n-1)!^2 (n-1)^{\frac{\nu -D}{\alpha D}} \,.
\end{align}
The Stirling approximation: $n!^2 \underset{n \rightarrow \infty}{\sim} 2 \pi n (n/e)^{2n}$ is used to obtain
\begin{equation}
N_{n \alpha} \underset{n \rightarrow \infty}{\sim} \lp\frac{\alpha^2 D}{r}\rp^{n-1} \lp \frac{n-1}{e} \rp^{2(n-1)} (n-1)^{\frac{\nu-D}{\alpha D}+1} \,.
\end{equation}
Switching back to $k=n\alpha$ leads to:
\begin{equation}
\label{eq_large_N_noise}
N_k \underset{k \rightarrow \infty}{\sim} \lp \sqrt{\frac{D}{r}}\frac{k}{e} \rp^\frac{2k}{\alpha} k^{\frac{2}{\alpha}\frac{\nu-D}{2D}-1} \,.
\end{equation}
The inverse Mellin transform of this moment is obtained in \cref{app_moments_large}.

The numerator of the general moments $M_{n \alpha}$ is given by 
\begin{align}
\prod_{j=1}^{n-1} \lbk \nu (j \alpha-1) + D j\alpha (j\alpha-1) \rbk &= (\alpha^2 D)^{n-1} (n-1)!^2 \prod_{j=1}^{n-1} \lp 1- \frac{1}{j \alpha} \rp \prod_{j=1}^{n-1} \lp 1 +\frac{\nu}{j \alpha D} \rp \\
& \underset{n \rightarrow \infty}{\sim} (\alpha^2 D)^{n-1} (n-1)!^2 (n-1)^{-\frac{1}{\alpha}} (n-1)^{\frac{\nu}{\alpha D}} \,,
\end{align}
which is identical to \cref{eq_large_N_num_noise}, so that the moment $M_k$ is equal to $N_k$ given by \cref{eq_large_N_noise}, and leads to the same distribution for large sizes.

\subsection{Small size limit}
\label{app_small_N}

In order to understand intuitively the case splitting into two regimes of the population distribution in the small size limit (\cref{eq_psi_small}), we give here an argument adapted from the fragmentation theory \cite{cheng_scaling_1988}, which is also easily generalizable to the lineage case. 
We multiply the steady-state \cref{eq_pop_full} by $x^k$ and integrate over $x$:
\begin{equation}
\label{eq_mom_pop_small}
r_0(m L_k -1) M_{k+\alpha_0} \sim \Lambda M_k - \nu_0 k M_{k+\beta_0-1} \,.
\end{equation}
From the power-law behavior of $b$ near $0$, we get that not all moments $L_k$ exists: there is a critical $k_c<0$ under which $L_k$ diverges. Thus, note that the equivalent sign `$\sim$' is used here to indicate the limit where $k$ is small enough to approximate the rates by their behavior near $x=0$, but still larger than $k_c$.

First we consider the case $\alpha_0 >0$. When letting $k \rightarrow k_c^+$, the left hand side of \cref{eq_mom_pop_small} diverges, and so must the moment $M$ of lowest order in the right hand side. When $\beta_0-1 \geq 0$, $M_k$ is the moment of lowest order and must diverge, while $M_{k+\beta_0-1}$ and $M_{k+\alpha_0}$ converge since they depend only on moments $L_k$ of order $k>k_c$. Therefore, $M_k \propto L_k$ where the proportionality constant is finite and positive, so that $\psi$ is given by the same power law as $b$: $\psi(x) \underset{x \rightarrow 0}{\sim} x^{\gamma_0}$. On the other hand, when $\beta_0-1 < 0$, the moment of lowest order is $M_{k+\beta_0-1}$ and thus $M_{k+\beta_0-1} \propto L_k$, where the proportionality constant is finite and positive, because $k_c<0$. In that case, Mellin transform properties tell us that $\psi(x) \underset{x \rightarrow 0}{\sim} x^{\gamma_0+1-\beta_0}$. When $\alpha_0=0$, the stability condition \cref{eq_cond_bet_alp_0} reads $\beta_0-1 < 0$, so that we are in the second case. 

The lineage equation on moments is similarly obtained by multiplying \cref{eq_lin_full} by $x^k$ and integrating over $x$:
\begin{equation}
\label{eq_mom_lin_small}
r_0(L_k -1) N_{k+\alpha_0} \sim - \nu_0 k N_{k+\beta_0-1} \,.
\end{equation}
The fundamental difference with the population case is the absence of terms in $N_k$. As a consequence, when $k \to k_c^+$ the left hand side of \cref{eq_mom_lin_small} diverges, so must $N_{k_c+\beta_0-1}$ regardless of the value of $\beta_0$. Let us explain this more precisely and show that $N_{k_c+\alpha_0}$ is non-diverging.
We define $\rho=\alpha_0-\beta_0+1 >0$, and $n$ such that $k=1-n\rho$, therefore \cref{eq_mom_lin_small} reads
\begin{equation}
N_{1-n \rho} \sim \frac{r_0}{\nu_0} \ \frac{1 - L_{1-n \rho + 1- \beta_0 } }{1-n \rho + 1- \beta_0 } N_{1-(n-1) \rho} \,.
\end{equation}
Iterating this relation leads to the general term:
\begin{equation}
\label{eq_rr_small}
N_{1-n \rho} \sim N_{1} \lp \frac{r_0}{\nu_0} \rp^{n} \prod_{j=1}^{n}  \ \frac{1 - L_{1-n \rho + 1- \beta_0 } }{1-n \rho + 1- \beta_0 } \,.
\end{equation}
Since all moments $L_k$ for $k>k_c$ exist, we get from this relation that $N_{k_c+\alpha_0}$ exists. Indeed, for $1-n \rho = k_c+\alpha_0$, the moment $L$ of lowest order in the product is of order $1-n \rho + 1- \beta_0=k_c +\alpha_0 + 1- \beta_0 > k_c$ because of \cref{eq_cond_bet_alp_0}, so that $N_{k_c+\alpha_0}$ converges. 
On the other hand, $N_{1-n \rho}$ diverges when $n$ is such that $1-n \rho + 1- \beta_0 = k_c$, which gives $N_{k_c+\beta_0-1} \propto L_{k_c}$ with a positive and finite proportionality constant, so that:
\begin{equation}
\phi(x) \underset{x \rightarrow 0}{\sim} x^{\gamma_0+1-\beta_0} \,.
\end{equation}

\section{Mellin transform of polynomial-exponential distribution}
\label{app_moments_large}

For a distribution $y$ characterized by its large $x$ behavior:
\begin{equation}
y(x) \underset{x \rightarrow \infty}{\sim} x^{\eta-\lambda(\mu-1/2)-1} \exp \lbk -\frac{x^\lambda}{\lambda \omega} \rbk \,,
\end{equation}
the moments of large order read
\begin{align}
M_k &\underset{k \rightarrow \infty}{\sim} \int_{0}^{\infty} \di x \ x^{k+\eta-\lambda(\mu-1/2)-1} e^{-\frac{x^\lambda}{\lambda \omega}} \\
&\sim \lambda^{-1} (\lambda \omega)^{(k+\eta)/\lambda -\mu+1/2}\int_{0}^{\infty} \di t \ t^{(k+\eta)/\lambda -\mu-1/2} e^{-t} \,,
\end{align}
where we went from the first to the second line using the change of variable $t=x^{\lambda}/\lambda \omega$. We recognize the function $\Gamma(z)=\int_{0}^{\infty} \di t \ t^{z-1} e^{-t}$ in the second line with $z=(k+\eta)/\lambda -\mu+1/2$, and we use the Stirling approximation: $\Gamma(z+1) \underset{z \rightarrow \infty}{\sim} \sqrt{2 \pi z} \lp \frac{z}{e} \rp^z$. Finally, the Mellin transform reads:
\begin{align}
M_k &\underset{k \rightarrow \infty}{\sim} \lambda^{-\frac{3}{2}} (\lambda \omega)^{(k+\eta)/\lambda -\mu+1/2} \sqrt{2 \pi (k+\eta-\lambda(\mu+1/2))} \lp \frac{k+\eta-\lambda(\mu+1/2)}{\lambda e} \rp^{\frac{k+\eta}{\lambda} -\mu-1/2}\\
&\sim \lp \frac{k\omega}{e} \rp^{\frac{k}{\lambda}} k^{\frac{\eta}{\lambda}-\mu} \,.
\end{align}
The large-size asymptotic behaviors of the distributions for noiseless (\cref{eq_lin_large}) and noisy (\cref{eq_noise_phi}) single-cell growth are then obtained from the moments of large orders \cref{eq_large_N} and \cref{eq_large_N_noise} with $\lambda=\alpha-\beta+1$, $\eta=1-\beta$, $\mu=1/2$, $\omega=\nu/r$ and $\lambda=\alpha/2$, $\eta=(\nu-D)/(2D)$, $\mu=1$, $\omega=\sqrt{D/r}$, respectively.

\end{document}